\documentclass[11pt,letterpaper]{article}

\usepackage[left=1in,right=1in,top=1in,bottom=0.9in]{geometry}

\usepackage{graphicx}
\usepackage[table,xcdraw]{xcolor}
\usepackage{wrapfig,lipsum,booktabs}
\usepackage{setspace}
\usepackage{tikz}
\usepackage{booktabs}
\usepackage{tabularx}
\usepackage{array}
\usetikzlibrary{decorations.shapes,shapes.geometric,shadows,arrows,automata,positioning,calendar,mindmap,backgrounds,scopes,chains,er,3d,matrix,fit}

\usepackage{wrapfig}

\usepackage[utf8]{inputenc}
\usepackage{amsmath}
\usepackage{amsfonts,times}
\usepackage{cite}

\usepackage{tikz,lipsum,lmodern}
\usepackage[most]{tcolorbox}

\usepackage{amssymb}
\usepackage{xurl}
\usepackage{comment}

\usepackage[compact]{titlesec}
    \titlespacing{\section}{0pt}{3ex}{2ex}
    \titlespacing{\subsection}{0pt}{2.0ex}{1.0ex}
    \titlespacing{\subsubsection}{0pt}{1.0ex}{0.5ex}

\usepackage{array}
\newcolumntype{L}{>{\centering\arraybackslash}m{3cm}}
\newcolumntype{Y}{>{\raggedright\arraybackslash}X}

\title{\bfseries The Internet of Agentic AI: Communication, Coordination, and Collective Intelligence at Scale}
\author{Quanyan Zhu\thanks{Quanyan Zhu is with the Department of Electrical and Computer Engineering and the NYU Center for Cybersecurity, NYU Tandon School of Engineering, New York University, Brooklyn, NY 11201, USA. Contact: \texttt{qz494@nyu.edu}.}}
\date{}

\begin{document}

\setstretch{1.67}

\pagenumbering{gobble}
\maketitle

\begin{abstract}
The rapid emergence of autonomous AI agents is transforming artificial intelligence from isolated model inference into distributed systems of reasoning, communication, and action. This paper develops the vision of the \emph{Internet of Agentic AI} (IoAI): an open ecosystem in which heterogeneous agents discover one another, negotiate responsibilities, exchange context, invoke tools, and execute workflows across cloud, edge, device, organizational, and cyber-physical environments. We synthesize foundations from single-agent agentic AI, multi-agent systems, distributed computing, communication networks, game theory, and security engineering to characterize the architectures and mechanisms required for scalable agent ecosystems. The paper examines agent deployment models, workflow lifecycles, communication protocols, interoperability layers, resource-management challenges, and trust architectures, with case studies in adaptive manufacturing and distributed operational coordination. The resulting framework highlights the central research challenges of controlled emergence, semantic interoperability, secure identity, incentive-compatible coordination, resource-aware orchestration, and governance for large-scale networks of autonomous agents.
\end{abstract}

\noindent\textbf{Keywords:} agentic AI; multi-agent systems; IoAI; distributed intelligence; agent communication protocols; interoperability; trust and security; resource management.



\section{Introduction}

Recent advances in artificial intelligence are driving a fundamental transition from isolated AI models toward large-scale ecosystems of interacting autonomous agents. Rather than relying on a single monolithic intelligence architecture, modern AI systems increasingly consist of collections of heterogeneous AI agents that communicate, coordinate, reason, and act collaboratively over distributed computational infrastructures. This emerging paradigm may be viewed as the \emph{Internet of Agentic AI} (IoAI), in which autonomous AI agents interact with one another and with external computational, physical, organizational, and cyber-physical systems through networked communication environments.

\begin{figure*}[t]
\centering
\includegraphics[width=0.92\textwidth]{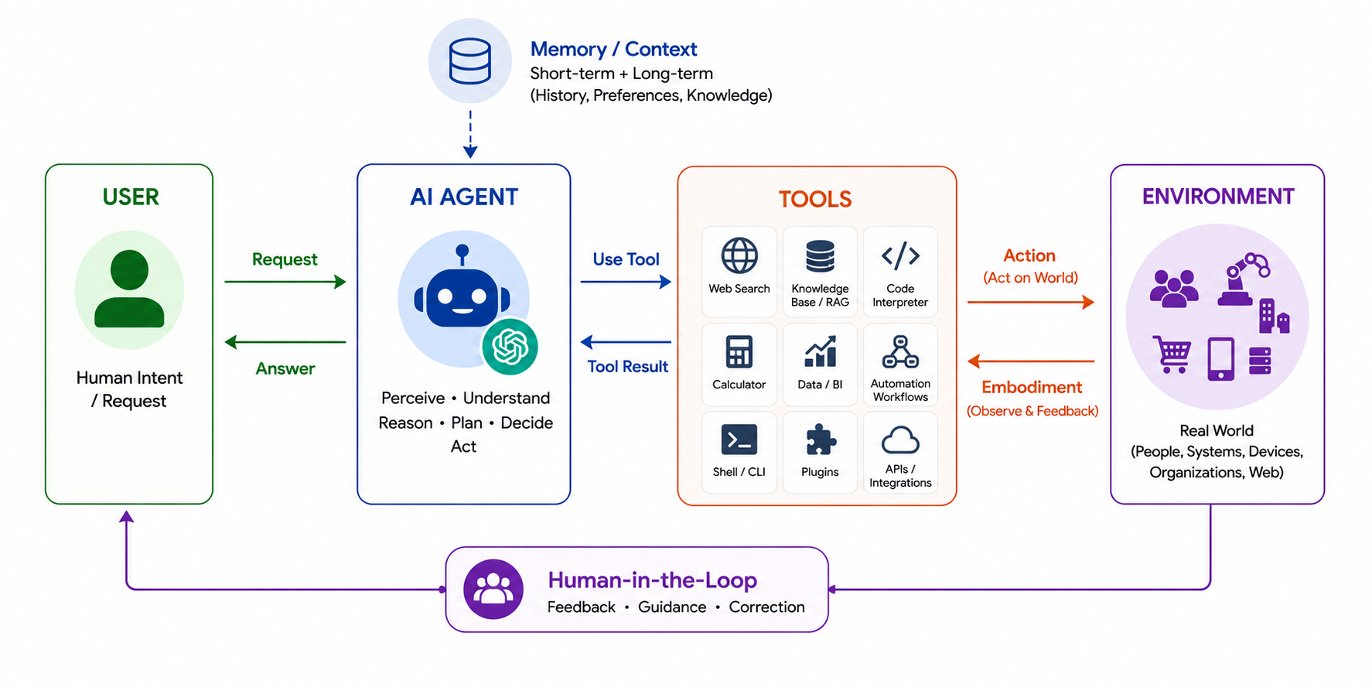}
\caption{
Conceptual architecture of an agentic AI system. The AI agent interacts with users, memory systems, external tools, and real-world environments through a closed-loop reasoning and action process. The agent receives user requests, performs contextual reasoning and planning, invokes external tools and APIs, executes actions in external environments, and incorporates feedback through observation and human-in-the-loop supervision. This architecture extends conventional AI systems beyond passive inference toward autonomous goal-directed behavior operating over computational and physical environments.
}
\label{fig:agentic_ai_architecture}
\end{figure*}

A fundamental building block underlying this paradigm is the notion of \emph{agentic AI}. Unlike conventional AI systems that primarily operate as passive inference engines responding to isolated user prompts, agentic AI systems are autonomous computational entities capable of perceiving contextual information, reasoning about objectives, planning actions, invoking external tools, interacting with environments, and adapting dynamically through feedback. Figure~\ref{fig:agentic_ai_architecture} illustrates the conceptual architecture of an agentic AI system. The AI agent interacts bidirectionally with users, external tools, memory systems, and real-world environments while operating within a human-in-the-loop feedback structure. Through access to external tools such as web search, retrieval systems, code interpreters, APIs, automation pipelines, and databases, the agent becomes capable of executing complex long-horizon tasks extending beyond conventional text generation.

The emergence of large language models, multimodal foundation models, retrieval-augmented reasoning systems, and tool-augmented AI architectures is accelerating the development of such agentic systems. AI agents are increasingly capable of invoking external APIs, interacting with databases and sensing infrastructures, generating executable code, coordinating with other agents, and recursively incorporating information received from distributed sources. Consequently, modern AI systems increasingly resemble decentralized societies of interacting reasoning agents rather than isolated machine-learning models.

\begin{figure*}[t]
\centering
\includegraphics[width=0.96\textwidth]{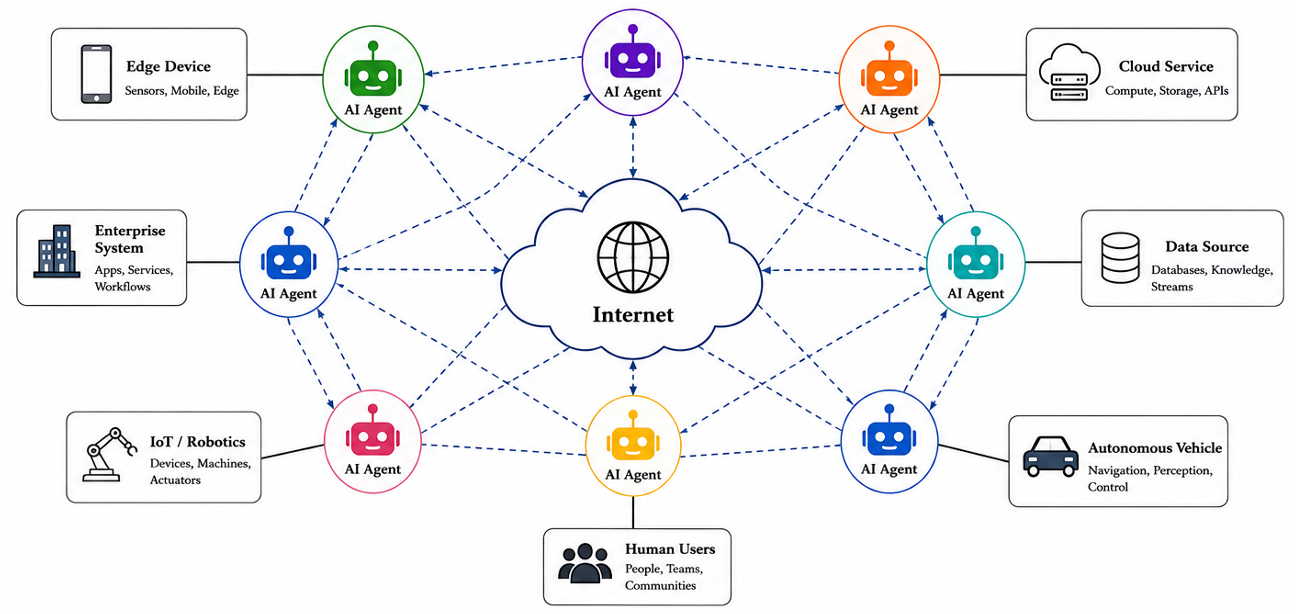}
\caption{
Conceptual illustration of IoAI. Autonomous AI agents communicate and coordinate over a distributed communication substrate while interacting with heterogeneous external systems including cloud services, enterprise infrastructures, edge devices, data repositories, robotic systems, autonomous vehicles, and human users. The architecture represents a decentralized ecosystem of interacting AI agents in which collective intelligence and system-level behavior emerge through recursive inter-agent communication, distributed reasoning, and collaborative workflow execution.
}
\label{fig:internet_agentic_ai}
\end{figure*}

As the number and sophistication of autonomous AI agents continue to increase, these systems naturally evolve toward networked ecosystems of interacting agents. Figure~\ref{fig:internet_agentic_ai} conceptually illustrates this broader IoAI vision. AI agents communicate and coordinate over a global communication substrate while interacting with heterogeneous environments including cloud services, enterprise systems, edge devices, robotic infrastructures, autonomous vehicles, data repositories, and human users. In contrast to traditional distributed computing systems, these agents are not passive computational endpoints executing fixed routines. Rather, they are autonomous reasoning entities capable of contextual inference, planning, communication, coordination, tool invocation, and adaptive decision-making. The resulting architecture forms a distributed cognitive ecosystem in which intelligence emerges collectively through recursive inter-agent interactions.

This transition is particularly important in large-scale operational environments in which no single AI agent possesses sufficient capabilities, information, computational resources, or situational awareness to execute complex tasks independently. Examples naturally arise in distributed cyber defense, collaborative scientific discovery, autonomous logistics coordination, intelligent sensing infrastructures, multi-robot systems, decentralized healthcare intelligence systems, smart manufacturing ecosystems, financial networks, and multi-agent software engineering environments. In such settings, complex objectives are decomposed into collections of interdependent subtasks executed collaboratively by specialized agents distributed across communication networks.

The growing importance of decentralized AI coordination is also increasingly recognized at the strategic level. Across scientific, industrial, healthcare, defense, and cyber-physical domains, future AI systems will need to support scalable, adaptive, and resilient collectives of heterogeneous agents capable of long-horizon coordination in dynamic environments. A central idea is \emph{controlled emergence}: the ability to harness useful system-level behavior arising from local agent interactions while preserving predictability, alignment, robustness, accountability, and operational coherence. This perspective does not assume that autonomy and control are opposites. Instead, it motivates architectures in which agents can self-organize, form temporary coalitions, and adapt to changing conditions while remaining bounded by communication protocols, trust mechanisms, resource constraints, and governance policies.

The architecture envisioned in such systems fundamentally differs from conventional centralized orchestration frameworks. Instead of relying on a single supervisory controller, decentralized AI ecosystems operate through local interactions among autonomous agents whose collective behavior emerges recursively across communication networks. This paradigm mirrors the principles underlying the scalability and resilience of the Internet itself, where robust global functionality emerges from decentralized local interaction rules rather than centralized global control. At the same time, agentic AI introduces additional layers of complexity because the networked entities are not passive endpoints: they reason, plan, invoke tools, negotiate responsibilities, and revise their behavior in response to new information.

These properties create scientific and engineering challenges that cannot be addressed by model capability alone. Workflow execution depends jointly on communication topology, capability distribution, resource allocation, information propagation, inter-agent coordination, trust, and recursive reasoning dynamics. AI agents may simultaneously participate in multiple workflows, compete for shared computational resources, exchange intermediate reasoning artifacts, influence one another through communication, and adapt dynamically to changing environmental conditions. As workflows propagate through the network, local reasoning decisions may amplify, attenuate, stabilize, or destabilize downstream system behavior. Consequently, IoAI should be studied as a distributed socio-technical and computational ecosystem rather than merely as a collection of stronger individual models.

This paper develops a vision and systems framework for IoAI. We synthesize foundations from agentic AI, multi-agent systems, distributed computing, communication networks, game theory, resource management, interoperability engineering, and security to characterize the architectures and mechanisms required for large-scale agent ecosystems. The scope of the paper is intentionally broad: it examines how agents are deployed across cloud, edge, device, organizational, and cyber-physical environments; how they discover one another and execute workflows; how communication protocols and interoperability layers enable collaboration; how computation and communication resources can be managed; and how trust, identity, security, and governance constrain autonomous coordination. The resulting framework identifies core design principles and research challenges for collective intelligence at scale, including controlled emergence, semantic interoperability, secure identity, incentive-compatible coordination, resource-aware orchestration, and trustworthy governance.

The remainder of the paper is organized as follows. Section~2 provides background on agentic AI and multi-agent systems, emphasizing how single-agent autonomy and bounded multi-agent coordination motivate Internet-scale agent ecosystems. Section~3 introduces IoAI as a distributed architecture, discusses illustrative scenarios, and describes agent deployment models and workflow lifecycles. Section~4 examines communication architectures, discovery services, naming, identity, messaging, agent communication protocols, and coordination mechanisms. Section~5 discusses interoperability, governance, and standardization as layered requirements for open agent ecosystems. Section~6 studies scalability and resource management across heterogeneous computing environments, and Section~7 analyzes trust and security challenges. Sections~8 and~9 present case studies in adaptive manufacturing and distributed operational coordination, respectively. Section~10 concludes with future research directions.

\section{Background}

Agentic artificial intelligence refers to AI systems capable of autonomously planning, coordinating, and executing multi-step workflows in pursuit of high-level goals. Unlike conventional generative AI systems that primarily respond to isolated prompts, agentic systems decompose objectives into sub-tasks, invoke external tools, maintain intermediate state, and adapt their plans as execution unfolds \cite{bandi2025rise,li2025agentic}. This shift from passive generation to autonomous action marks an important transition in the design of AI systems, particularly as large language models are increasingly combined with software tools, memory modules, retrieval systems, code execution environments, and application programming interfaces.

The rapid emergence of frameworks such as AutoGPT, BabyAGI, LangChain, LangGraph, CrewAI, AutoGen, Semantic Kernel, MetaGPT, CAMEL, and related orchestration platforms has accelerated the development of agentic workflows. These frameworks demonstrate how language models can function not merely as text generators, but as coordinators of modular computational processes involving planning, delegation, execution, and verification \cite{li2025agentic,langchain2026langgraph,microsoft2026agentframework,hong2023metagpt,li2023camel}. Industry platforms have further reinforced this trajectory through agent-oriented APIs and integrated tool ecosystems that enable agents to browse, compute, retrieve files, and interact with external services \cite{openai2025agents,openai2025chatgptagent}.

A broader vision now emerging is IoAI, in which heterogeneous AI agents are distributed across cloud, edge, and device environments and coordinate through shared communication protocols. In this model, agents may locate suitable collaborators, advertise capabilities, form coalitions, exchange intermediate context, and jointly execute workflows under incentive-compatible mechanisms \cite{yang2026internet}. Such a vision extends agentic AI from isolated software assistants to networked socio-technical systems, where interoperability, trust, communication standards, and economic incentives become central design concerns.

Recent work by Zhu and collaborators further highlights that LLM agents should be understood as strategic, social, and risk-bearing entities rather than only as workflow automation components. This line of work studies social cognition in LLM agents, LLM-mediated game-theoretic reasoning, cyber-resilient agentic workflows, privacy leakage from generative AI agents, and insurance mechanisms for agentic-AI risk \cite{liu2025prosocial,zhu2025llmstackelberg,zhu2025gametheoryllm,li2025cyberresilience,yang2026dpagents,zhu2026insurance}. These perspectives reinforce the need to study communication, incentives, privacy, security, and governance as first-class components of IoAI.

The development of distributed agentic systems also raises significant security and governance challenges. Autonomous agents require reliable mechanisms for identity, authentication, authorization, provenance, and revocation. Existing human-centric identity and access management systems are insufficient for ephemeral and autonomous agents that may act across organizational boundaries \cite{huang2025zerotrust}. Proposed approaches include public key infrastructure, decentralized identifiers, verifiable credentials, agent naming services, secure communication channels, and fine-grained access control \cite{huang2025zerotrust,keyfactor2025pki,silwer2025trust}. These mechanisms are necessary to mitigate risks such as impersonation, Sybil attacks, registry poisoning, unauthorized delegation, and cascading failures in multi-agent workflows.

Scalability and performance are equally important. Agentic workflows often involve multiple model invocations, external API calls, tool executions, and inter-agent communications, producing nontrivial latency and resource-management challenges. Recent work argues that efficient agentic AI will require heterogeneous computing architectures capable of mapping different workflow components to appropriate hardware resources, including GPUs, CPUs, memory-optimized systems, and edge devices \cite{wang2025heterogeneous}. Parallel execution and low-latency interconnects may partially offset coordination overhead, but the design of scalable orchestration mechanisms remains an open research problem \cite{bourgoyne2026dgx}.

Government and standards-oriented initiatives further indicate the strategic importance of this field. DARPA's MATHBAC program emphasizes the need for formal mathematical foundations for agent communication, including common protocols and generalizable principles for how AI agents collaborate, share information, and improve over time \cite{darpa2026mathbac}. These efforts suggest that agentic AI is no longer merely an application-layer engineering trend, but an emerging research area at the intersection of artificial intelligence, distributed systems, cybersecurity, communication networks, economics, and governance.

\subsection{Single-Agent Agentic AI: Foundations and Recent Advances}

Agentic AI represents a shift from passive generative models toward autonomous systems capable of perceiving their environment, reasoning about observations, and taking actions to achieve goals \cite{bandi2025rise,li2025agentic}. While the concept of intelligent agents has long been studied in artificial intelligence, recent advances in large language models (LLMs) have dramatically expanded their capabilities, enabling agents to operate in open-ended environments, interact with external tools, and pursue objectives over extended time horizons.

Traditional agent architectures, such as rule-based systems and Belief--Desire--Intention (BDI) agents, relied on symbolic reasoning, handcrafted knowledge representations, and predefined action policies \cite{wooldridge2009introduction,russell2021artificial}. Although effective in structured environments, these approaches often struggled with uncertainty, incomplete information, and rapidly changing conditions. Modern agentic systems instead leverage LLMs as flexible reasoning engines capable of interpreting instructions, generating plans, adapting to new information, and coordinating sequences of actions \cite{bandi2025rise,li2025agentic,zhu2025gametheoryllm}.

A distinguishing characteristic of contemporary agentic AI is the integration of memory, planning, and tool use within a unified decision-making framework. Modern agents maintain contextual memory across interactions, decompose complex objectives into subtasks, and invoke external resources such as search engines, databases, APIs, software tools, and simulation environments to accomplish goals \cite{park2023generative,openai2025agents}. Rather than producing a single response to a prompt, they operate through iterative perception--reasoning--action loops that support long-horizon decision making and adaptive behavior.

Because LLM agents reason through language, their behavior also reflects social, strategic, and cognitive dimensions that are less prominent in conventional software agents. Recent studies show that LLM agents can exhibit prosocial decision patterns under uncertainty and can participate in structured strategic interactions when game-theoretic reasoning is embedded into prompts and workflows \cite{liu2025prosocial,zhu2025llmstackelberg}. These findings suggest that single-agent autonomy cannot be evaluated only by task completion; it must also be evaluated through reliability, strategic behavior, privacy, and alignment under interaction.

Recent frameworks such as AutoGPT, BabyAGI, LangChain, LangGraph, CrewAI, AutoGen, Semantic Kernel, and OpenAI's Agents SDK have accelerated the development of agentic systems by providing reusable abstractions for memory management, workflow orchestration, planning, and tool integration \cite{li2025agentic,microsoft2026agentframework,openai2025agents}. These frameworks have transformed LLMs from standalone conversational models into autonomous software entities capable of executing multi-step workflows with limited human supervision. They also make agentic design more modular: memory, tool access, routing, evaluation, and human approval can be treated as configurable components rather than as ad hoc prompt-engineering patterns.

Despite these advances, single-agent systems face inherent limitations. As tasks grow in complexity, a single agent may become constrained by limited situational awareness, computational bottlenecks, or the need for diverse expertise. Long reasoning chains can accumulate errors, and a single agent may struggle to simultaneously perform planning, execution, monitoring, and adaptation. These challenges motivate the transition toward multi-agent and distributed agentic systems, where specialized agents collaborate to solve problems that exceed the capabilities of any individual agent \cite{li2025agentic,yang2026internet}. Consequently, modern research is increasingly focused on understanding how collections of autonomous agents can coordinate, communicate, and self-organize to achieve system-level intelligence.

\subsection{Multi-Agent Systems}

While single-agent systems provide a foundation for autonomous reasoning and action, many real-world problems require coordination among multiple decision-making entities. A \emph{multi-agent system} (MAS) consists of two or more autonomous agents that interact within a shared environment \cite{wooldridge2009introduction,stone2000multiagent}. Each agent may maintain private state, local observations, objectives, and decision policies, while coordinating with others through direct communication or indirectly by acting on the environment.

A MAS differs from a conventional distributed software system in that its components are not merely executing fixed protocols. Agents are autonomous, adaptive, and goal-directed. In modern LLM-based MAS, agents may communicate through flexible natural-language or semi-structured messages, negotiate task assignments, exchange knowledge, and revise their behavior as conditions change \cite{li2025agentic,bandi2025rise,li2023camel,liu2025prosocial}. Coordination therefore emerges not only from engineered workflows, but also from the reasoning, communication, and adaptation of the participating agents.

Multi-agent interactions range from fully cooperative to competitive. In cooperative systems, agents share a common objective and collaborate to improve collective performance, as in robotic teams, distributed sensing systems, or workflow automation platforms \cite{panait2005cooperative}. In competitive or mixed-motive systems, agents may have partially aligned or conflicting objectives, as in markets, cybersecurity, resource allocation, and strategic decision-making \cite{basar1999dynamic,shoham2008multiagent,zhu2025llmstackelberg}. Recent work on multilevel interactive equilibrium further suggests that equilibria among intelligent agents may need to account for internal computation, cognitive representations, and adaptive reasoning rather than only observable strategies \cite{chen2026mie}. This diversity makes MAS a natural framework for studying coordination, negotiation, coalition formation, and conflict resolution.

A defining feature of MAS is the distribution of information and decision authority. No agent necessarily has complete knowledge of the global system state. Instead, agents act on local observations and partial information while updating their behavior through interaction with others. This decentralization can improve scalability, robustness, and flexibility relative to centralized control, but it also introduces challenges such as communication overhead, coordination failures, trust management, and undesirable emergent behavior \cite{wooldridge2009introduction}.

Recent agentic AI architectures increasingly exploit specialization. Rather than assigning the same role to every agent, systems may include planning agents, execution agents, monitoring agents, retrieval agents, verification agents, or quality-assurance agents \cite{li2025agentic}. Complex objectives can then be decomposed into subtasks and distributed across agents with complementary capabilities. Frameworks such as AutoGen, CrewAI, LangGraph, MetaGPT, and Microsoft Agent Framework support this role-based organization, enabling groups of agents to execute workflows that exceed the capability of any individual participant \cite{microsoft2026agentframework,li2025agentic,hong2023metagpt}.

For example, in a manufacturing setting, a planning agent may generate production schedules, robotic agents may perform machining and assembly, inspection agents may monitor quality, and transport agents may move materials among workstations. No single agent controls the entire production process. Instead, performance emerges from the coordination of specialized agents that exchange information, respond to local conditions, and adapt their actions over time. Similar patterns arise in software engineering, scientific discovery, cyber defense, and logistics.

Thus, MAS provide the conceptual bridge between single-agent autonomy and large-scale networked agent ecosystems. They introduce the core mechanisms of collective intelligence: distributed knowledge, role specialization, communication, cooperation, competition, and emergent system behavior. These mechanisms become foundational for understanding IoAI, where coordination extends beyond a bounded team of agents to open, heterogeneous, and dynamically evolving agent networks.

\section{IoAI: Architecture, Deployment, and Workflows}

The evolution of artificial intelligence has progressed from standalone intelligent agents to collaborative multi-agent systems capable of distributed decision making. While multi-agent systems enable coordination among a bounded collection of agents operating within a common environment, many emerging applications require collaboration across organizational, geographic, and technological boundaries. Scientific discovery, advanced manufacturing, healthcare, cyber defense, logistics, and critical infrastructure management increasingly depend on expertise, resources, and data distributed across multiple institutions. These challenges motivate the concept of IoAI, a globally interconnected ecosystem in which autonomous agents identify collaborators, exchange information, negotiate responsibilities, share resources, and collectively execute workflows through standardized communication and trust mechanisms \cite{yang2026internet}.

IoAI extends the role of communication networks beyond information exchange to intelligence exchange. Just as the Internet transformed isolated computers into a global information infrastructure, IoAI seeks to transform isolated agents and bounded multi-agent systems into a distributed ecosystem of interoperable intelligence. In this vision, agents are no longer confined to a single application, organization, or computational platform. Instead, they become autonomous participants in a global network capable of advertising capabilities, requesting services, discovering collaborators, and forming dynamic coalitions to solve problems that exceed the capabilities of any individual agent or organization \cite{yang2026internet,li2025agentic}.

\begin{figure}[t]
    \centering
    \includegraphics[width=\textwidth]{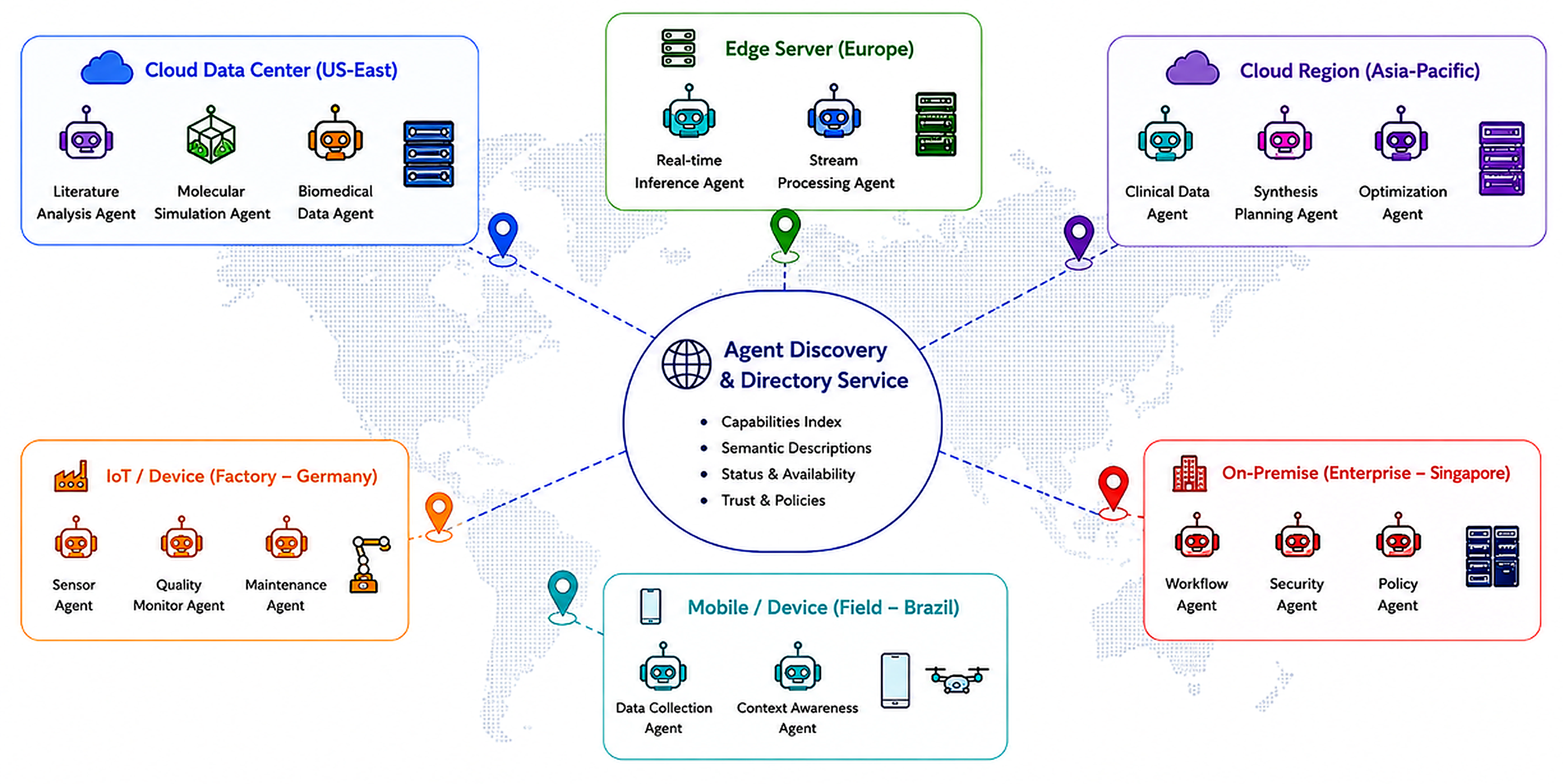}
    \caption{Distributed agent deployment in IoAI. Specialized agents operate across cloud data centers, edge servers, mobile and IoT environments, enterprise systems, and regional cloud platforms. A discovery and directory service supports capability indexing, semantic descriptions, availability information, and trust policies, enabling agents to find collaborators across heterogeneous infrastructures.}
    \label{fig:network_agent_deployment}
\end{figure}

Figure~\ref{fig:network_agent_deployment} gives a network-level view of this ecosystem. Agents may reside in cloud data centers, regional cloud platforms, edge servers, on-premise enterprise systems, mobile environments, and IoT devices. A shared agent discovery and directory service provides capability indexing, semantic descriptions, status and availability information, and trust-policy metadata that allow agents to locate appropriate collaborators across geographic and organizational boundaries. This architecture emphasizes that IoAI is not a single platform, but a coordination layer spanning many administrative domains and computational environments.

\subsection{Motivation and Benefits}

The primary motivation for IoAI is that many real-world challenges require capabilities that are inherently distributed. Modern problems often involve heterogeneous expertise, geographically dispersed resources, dynamically changing environments, and complex interdependencies among stakeholders. No single agent possesses all the knowledge, computational power, sensing capabilities, or authority required to solve such problems independently. Consequently, intelligence itself must become distributed.

Table~\ref{tab:ioai_benefits} summarizes the core benefits that follow from this distributed view of intelligence. These benefits provide a common vocabulary for the rest of the paper: IoAI is valuable not merely because many agents communicate, but because communication, coordination, and governance allow agent networks to execute missions, scale elastically, adapt to changing conditions, remain resilient under disruption, and produce controlled forms of collective behavior.

\begin{table}[t]
\centering
\caption{Core benefits of the Internet of Agentic AI (IoAI).}
\label{tab:ioai_benefits}
\small
\setlength{\tabcolsep}{4.5pt}
\renewcommand{\arraystretch}{1.16}
\begin{tabularx}{\textwidth}{@{}>{\raggedright\arraybackslash}p{3.1cm}Y@{}}
\toprule
\textbf{Core benefit} & \textbf{Description} \\
\midrule
\textbf{Collective Intelligence} & Multiple agents combine knowledge, reasoning, and expertise to achieve outcomes beyond the capability of any individual agent. \\
\addlinespace[2pt]
\textbf{Mission Execution} & Distributed agents coordinate planning, reasoning, and action to accomplish complex, long-horizon, multi-stage objectives that exceed the capacity of a single agent. \\
\addlinespace[2pt]
\textbf{Elastic Scalability} & Intelligence can be expanded through the addition and coordination of agents rather than by continuously increasing the size of individual models. \\
\addlinespace[2pt]
\textbf{Adaptive Coordination} & Agents dynamically reorganize tasks, workflows, and collaborations in response to changing conditions and objectives. \\
\addlinespace[2pt]
\textbf{System Resilience} & The system maintains and recovers functionality under failures, attacks, uncertainty, and disruptions through redundancy, self-healing, and adaptive reconfiguration. \\
\addlinespace[2pt]
\textbf{Controlled Emergence} & Novel system-level capabilities, behaviors, and insights arise from interactions among agents, exceeding what is explicitly designed into any individual agent while remaining aligned with system objectives. \\
\bottomrule
\end{tabularx}
\end{table}

Collective intelligence and mission execution are the most direct benefits of IoAI. Individual agents are constrained by local information, limited computational resources, specialized expertise, and bounded authority. Through large-scale interaction, however, agents can combine partial knowledge and complementary capabilities into coordinated workflows that achieve objectives no single participant could complete alone. Similar phenomena are observed in biological systems, social organizations, and distributed computing networks, where global functionality emerges from local interactions among autonomous entities \cite{camazine2003self,holland2014complexity}. IoAI extends this principle to artificial intelligence by enabling distributed agents to collectively reason, learn, plan, and act.

Elastic scalability follows from distributing intelligence across cloud infrastructures, edge devices, robots, sensors, autonomous vehicles, and mobile platforms, as illustrated in Figure~\ref{fig:network_agent_deployment}. Centralized AI systems concentrate computation and decision making within a limited set of resources, creating bottlenecks as workloads increase. In contrast, IoAI allows new agents, tools, data services, and computational resources to join workflows as demand grows. Scalability therefore comes not only from larger models or larger data centers, but also from the ability to coordinate many specialized agents across heterogeneous environments \cite{varghese2018fog,yang2026internet}.

Adaptive coordination and system resilience are essential because real-world environments rarely remain stable. Centralized architectures often contain single points of failure whose compromise can disrupt entire operations. Distributed agent ecosystems can continue functioning despite failures, communication disruptions, cyber attacks, resource shortages, or changes in mission priority. Tasks can be reassigned, alternative coalitions can form, and resources can be dynamically reallocated. As a result, system performance degrades gracefully rather than catastrophically when disruptions occur \cite{olfati2007consensus,ren2007information}.

Controlled emergence captures a central design tension in IoAI. Rather than constructing monolithic AI systems capable of performing every function, IoAI allows agents to specialize in planning, optimization, sensing, verification, execution, negotiation, monitoring, or coordination. Complex objectives can then be decomposed and distributed among agents possessing complementary expertise, increasing efficiency and adaptability \cite{wooldridge2009introduction,shoham2008multiagent}. At the same time, emergent behavior must remain aligned with system objectives. IoAI therefore requires protocols, incentives, trust mechanisms, and monitoring structures that allow useful system-level capabilities to arise while bounding unsafe, unstable, or strategically manipulative behaviors.

\subsection{Illustrative Scenarios}

The value of IoAI becomes particularly apparent when examining realistic application scenarios. Consider a healthcare ecosystem involving hospitals, imaging centers, insurance providers, specialists, and telemedicine services. A diagnostic agent may identify a patient requiring specialized treatment but lack access to the necessary expertise or resources. Through IoAI, the diagnostic agent can discover radiology agents, insurance-verification agents, scheduling agents, and specialist-consultation agents distributed across multiple organizations. These agents negotiate responsibilities, form a temporary coalition, and execute a coordinated treatment workflow. The resulting service emerges from the collective capabilities of the participating agents rather than from any individual institution.

\begin{figure}[t]
    \centering
    \includegraphics[width=\textwidth]{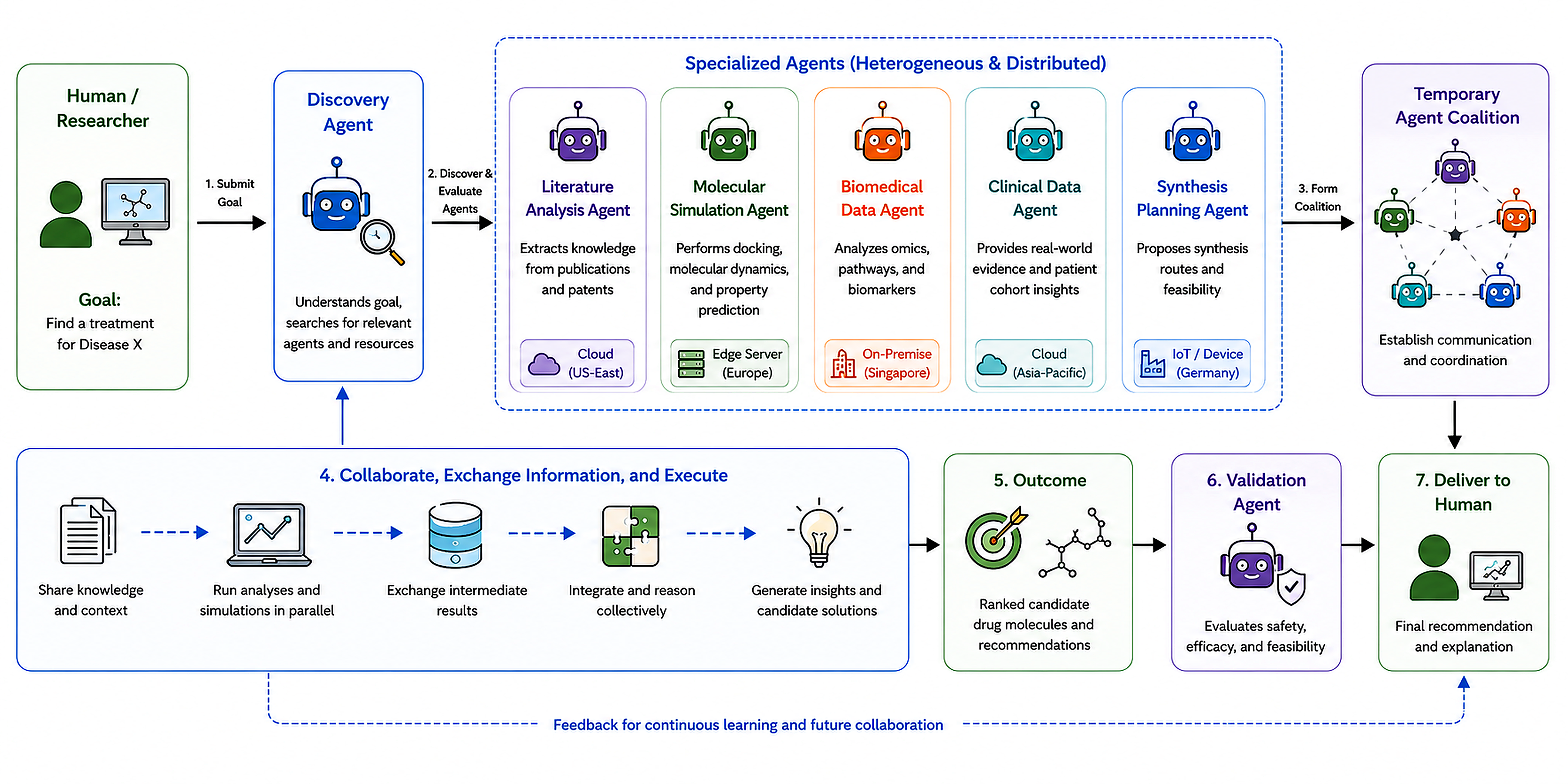}
    \caption{Collaborative drug discovery in IoAI. A human researcher submits a scientific goal, a discovery agent identifies specialized agents across organizations, and the selected agents form a temporary coalition to exchange knowledge, execute analyses, integrate intermediate results, validate candidate solutions, and deliver recommendations. The example illustrates cross-organizational agent discovery, coalition formation, workflow execution, validation, and feedback.}
    \label{fig:collaborative_drug_discovery}
\end{figure}

Figure~\ref{fig:collaborative_drug_discovery} illustrates a closely related scientific scenario that builds on the distributed infrastructure in Figure~\ref{fig:network_agent_deployment}: collaborative drug discovery across organizational boundaries. A human researcher submits the goal of finding a treatment for a disease, after which a discovery agent identifies specialized agents distributed across universities, hospitals, research institutes, high-performance computing centers, and pharmaceutical organizations. These agents contribute complementary capabilities, including literature analysis, molecular simulation, biomedical-data analysis, clinical-data interpretation, synthesis planning, and validation. By locating complementary agents, forming a temporary coalition, exchanging intermediate results, running analyses in parallel, and integrating evidence, the agents can produce candidate drug molecules and recommendations beyond the capability of any single agent or institution.

A similar pattern arises in scientific discovery. Modern scientific workflows frequently involve experimental laboratories, high-performance computing centers, cloud infrastructures, simulation environments, and data repositories distributed around the world. Federated agents can coordinate experiments, simulations, and analyses across these infrastructures, dynamically adapting to changing resource availability and experimental outcomes. Such workflows enable levels of scientific coordination that would be difficult to achieve through traditional centralized orchestration.

In advanced manufacturing, intelligent agents representing machines, robots, quality-control systems, logistics platforms, and human supervisors can dynamically form task-oriented coalitions. When a machine fails, maintenance agents can coordinate repairs while production agents reorganize schedules and logistics agents reroute materials. The factory effectively becomes a self-organizing ecosystem of collaborating agents capable of adapting to disruptions in real time. Across these scenarios, collaboration among distributed agents produces capabilities that do not reside in any individual participant: the value of the network comes from matching capabilities, negotiating responsibilities, and composing partial contributions into coordinated workflows.

\subsection{Agent Types, Deployment Models, and Distributed Intelligence}

A defining characteristic of IoAI is the diversity of participating agents and computational environments. Unlike traditional multi-agent systems, which are often deployed within a single computational infrastructure, Internet-scale agent ecosystems span cloud platforms, edge devices, scientific computing facilities, industrial systems, mobile environments, and embedded platforms. Consequently, understanding IoAI requires considering not only what agents do, but also where they execute and how computational resources are distributed across the network \cite{yang2026internet,varghese2018fog}. 

From a functional perspective, agents may specialize in planning, optimization, sensing, verification, execution, monitoring, negotiation, coordination, or decision support. Such specialization enables complex objectives to be decomposed into smaller subtasks that can be assigned to agents possessing complementary expertise. This division of labor allows distributed agent ecosystems to solve problems whose scale and complexity exceed the capabilities of individual agents \cite{wooldridge2009introduction,li2025agentic}.

Beyond functional specialization, agent ecosystems are characterized by diverse deployment models. The choice of deployment architecture influences system performance, scalability, latency, resilience, privacy, and operational cost. The deployment architecture introduced in Figure~\ref{fig:network_agent_deployment} makes this distribution explicit, while Table~\ref{tab:deployment_models} summarizes the main deployment models and the system-design trade-offs that they introduce. These deployment choices shape not only where computation occurs, but also how quickly agents can respond, which data they may access, and which governance rules apply.

\begin{table}[t]
\centering
\caption{Representative deployment models in IoAI.}
\label{tab:deployment_models}
\small
\setlength{\tabcolsep}{3.5pt}
\renewcommand{\arraystretch}{1.15}
\begin{tabularx}{\textwidth}{@{}>{\raggedright\arraybackslash}p{2.25cm}YYY@{}}
\toprule
\textbf{Model} & \textbf{Typical agents} & \textbf{Primary strengths} & \textbf{Main constraints} \\
\midrule
Cloud & Large-model planners, enterprise workflow agents, simulation coordinators & Elastic compute, centralized observability, access to large foundation models and data services & Latency, data-governance exposure, dependence on provider availability \\
\addlinespace[2pt]
Edge/fog & Industrial controllers, robotic agents, local sensing and filtering agents & Low latency, bandwidth reduction, resilience under intermittent connectivity & Limited compute, heterogeneous hardware, local security management \\
\addlinespace[2pt]
On-device & Personal assistants, wearables, drones, mobile agents & Privacy, offline operation, immediate responsiveness & Energy, memory, model size, physical compromise risk \\
\addlinespace[2pt]
Federated & Cross-institution scientific, healthcare, manufacturing, or infrastructure agents & Preserves local autonomy while enabling shared workflows across organizations \cite{pauloski2025academy} & Policy heterogeneity, identity federation, auditability, incentive alignment \\
\addlinespace[2pt]
Elastic & Ephemeral task agents, bursty workflow workers, event-driven tool agents & Elastic scaling, cost efficiency, rapid instantiation for short-lived tasks & Cold-start latency, state management, provenance tracking \\
\addlinespace[2pt]
Economies & Service-provider agents, compute-market agents, data-product agents & Dynamic resource discovery, specialization, market-based allocation & Trust, pricing, settlement, liability, and strategic manipulation \\
\bottomrule
\end{tabularx}
\end{table}

\subsubsection{Cloud Agents}

Many contemporary agentic systems are deployed in cloud data centers, where agents execute within virtual machines, containers, or serverless infrastructures. Cloud deployment provides access to large-scale computational resources, extensive storage capabilities, and high-performance networking. Such environments are particularly well suited for computationally intensive tasks involving large foundation models, long-horizon planning, and large-scale analytics. For example, a scientific research agent may employ large language models to synthesize literature, generate hypotheses, and coordinate simulations that require thousands of GPU-hours. Similarly, enterprise planning agents may optimize global supply chains by analyzing vast quantities of operational data. Cloud infrastructures make such capabilities feasible because they aggregate substantial computational resources in centralized locations. However, cloud-centric architectures also introduce communication latency, privacy concerns, and potential dependence on particular cloud providers. Consequently, cloud agents are increasingly complemented by edge and local intelligence, especially when decisions must be made near data sources or under strict governance constraints.

\subsubsection{Edge and Fog Agents}

At the opposite end of the deployment spectrum are \emph{edge agents}, which operate close to data sources such as sensors, robots, industrial equipment, autonomous vehicles, and Internet-of-Things devices. Edge agents perform local sensing, reasoning, and control, reducing communication latency and minimizing bandwidth requirements. Consider an autonomous manufacturing facility: vision agents inspecting products on an assembly line cannot afford to wait for cloud-based responses because production decisions must be made within milliseconds. Similarly, autonomous vehicles require local reasoning capabilities to maintain safe operation even when connectivity is intermittent. Edge agents therefore provide low-latency intelligence directly where decisions must be made, while still allowing higher-level cloud agents to contribute planning, aggregation, and long-horizon optimization.

Fog-computing architectures extend this concept by introducing intermediate computational layers between edge devices and cloud platforms. Rather than viewing cloud and edge computing as separate paradigms, fog computing creates a continuum of distributed intelligence in which computational tasks migrate dynamically among devices, local servers, and cloud infrastructures according to workload demands and latency requirements \cite{varghese2018fog}. This continuum is especially important for agent ecosystems because reasoning, sensing, control, and coordination may each have different latency, privacy, and resource requirements.

\subsubsection{On-Device Agents}

Recent advances in lightweight language models, efficient inference techniques, and specialized AI accelerators have enabled autonomous agents to operate directly on mobile devices, wearables, drones, and embedded systems. These \emph{on-device agents} offer immediate responsiveness, offline operation, and enhanced privacy because sensitive data need not leave the local device. For example, a personal health assistant running on a wearable device may continuously monitor physiological signals, detect anomalies, and provide recommendations without transmitting private medical data to external servers. Similarly, drones operating in contested environments may continue functioning despite losing connectivity to cloud infrastructure. Such capabilities are becoming increasingly feasible as hardware accelerators and model compression techniques improve. Nevertheless, on-device agents remain constrained by limited computational resources, energy availability, memory capacity, and physical security risks, which means they often participate as local decision nodes within broader cloud-edge-device workflows rather than as fully self-sufficient agents.

\subsubsection{Federated Agents}

Many real-world applications require collaboration across organizational boundaries. \emph{Federated agents} operate in such environments by orchestrating resources distributed across multiple institutions while preserving local autonomy. Scientific discovery provides an illustrative example. A research workflow may involve laboratory experiments at one institution, large-scale simulations at a supercomputing center, machine-learning analysis in a cloud environment, and data repositories located elsewhere. Federated agents coordinate these distributed resources while respecting local governance policies and ownership constraints; recent scientific-workflow middleware demonstrates how agents can be deployed across HPC systems, experimental facilities, and data repositories \cite{pauloski2025academy}. Similar patterns arise in healthcare systems, manufacturing supply chains, and critical infrastructure operations. Rather than requiring centralized ownership, federated agents enable independent organizations to contribute resources while maintaining control over their own assets and policies, making them central to the cross-institutional coalition formation envisioned by IoAI.

\subsubsection{Hybrid and Serverless Agents}

Not all agents require persistent execution. Some tasks arise sporadically and can be handled by agents instantiated on demand. Serverless computing platforms and container orchestration systems increasingly support such \emph{ephemeral agents}. In these architectures, agents are dynamically created when tasks arrive and terminated after completion. This model provides elastic scalability and efficient resource utilization because computational resources are consumed only when needed. Hybrid deployments frequently combine persistent cloud agents, edge agents, and ephemeral serverless agents to balance responsiveness, scalability, and operational cost. The resulting architecture allows long-lived agents to maintain context and policy continuity while short-lived agents absorb bursty workloads or specialized subtasks.

\subsubsection{Agent Economies and Distributed Compute}

An increasingly important vision for IoAI is the emergence of distributed agent economies. In such environments, agents act both as consumers and providers of services. Agents may advertise computational resources, specialized expertise, data products, simulation capabilities, or analytical services. Other agents can discover these resources, negotiate usage agreements, and incorporate them into larger workflows. This economic framing connects technical interoperability with incentives, since agents must have reasons to provide reliable services and mechanisms for pricing, reputation, and accountability.

More ambitious proposals envision distributed compute markets in which agents leverage idle computational resources contributed by participants across the network. Similar to volunteer computing, federated learning, and decentralized blockchain infrastructures, such ecosystems could dynamically allocate workloads according to cost, availability, trust, and performance considerations. Although still experimental, distributed compute markets represent a potentially transformative mechanism for scaling future agent ecosystems \cite{yang2026internet}. These deployment models span a continuum from embedded devices and edge platforms to cloud data centers and globally distributed infrastructures. Integrating heterogeneous resources into a coherent ecosystem of autonomous agents is a key source of scalability, adaptability, and resilience because it allows workflows to place intelligence where it is most useful rather than where a monolithic system happens to be deployed \cite{yang2026internet}.

\subsection{Agentic Workflows in IoAI}

IoAI is fundamentally a workflow-driven ecosystem in which autonomous agents cooperate on tasks that require distributed capabilities. Unlike traditional software workflows, where interactions are predefined and centrally orchestrated, agentic workflows are dynamic, adaptive, and often decentralized. Agents must discover collaborators, negotiate responsibilities, allocate tasks, execute actions, monitor progress, and adapt to changing conditions throughout the lifetime of a mission \cite{yang2026internet,li2025agentic}.

The lifecycle of a typical agentic workflow consists of four interconnected phases: discovery and negotiation, task allocation and delegation, execution and monitoring, and adaptation and composition. The collaborative drug-discovery example in Figure~\ref{fig:collaborative_drug_discovery} provides a concrete instance of this lifecycle, while Figure~\ref{fig:network_agent_deployment} shows the discovery substrate that makes such cross-organizational workflows possible. The discovery agent first searches for complementary capabilities, the participating organizations negotiate a temporary coalition, specialized agents execute literature analysis, simulation, clinical-data analysis, synthesis planning, and validation, and the resulting feedback loop supports continuous learning and future collaboration. The same lifecycle generalizes to other domains: agents must first find each other, then establish the terms of collaboration, then execute interdependent tasks, and finally revise the coalition as new evidence, constraints, or opportunities appear.

\subsubsection{Discovery and Negotiation}

The workflow begins when an agent identifies a goal that cannot be completed independently. The agent must first discover collaborators possessing the required capabilities. Discovery may occur through centralized registries, distributed directories, peer-to-peer broadcasts, or semantic capability repositories, including the kind of directory service depicted in Figure~\ref{fig:network_agent_deployment}. Agents typically advertise capabilities, resources, service endpoints, and operational constraints through machine-readable metadata. Once candidate collaborators are identified, negotiation mechanisms determine how responsibilities will be distributed. Agents may exchange bids, requests, commitments, and service guarantees. Classical coordination mechanisms such as auctions and Contract-Net protocols remain relevant, while LLM-enabled agents can also engage in richer natural-language negotiations that clarify assumptions, constraints, and expected outputs before execution begins \cite{wooldridge2009introduction}.

\subsubsection{Task Allocation and Delegation}

Following negotiation, tasks are allocated among participating agents. Complex objectives are decomposed into smaller subtasks that can be executed independently or in parallel. Delegation mechanisms specify responsibilities, deadlines, permissions, and dependencies among tasks \cite{li2025agentic}. For example, a healthcare workflow may decompose patient diagnosis, treatment planning, insurance verification, and scheduling among specialized agents. Similarly, a manufacturing workflow may allocate machining, transportation, assembly, and quality-control tasks to different agents possessing complementary capabilities. Effective task allocation must therefore account not only for functional expertise, but also for availability, trust, latency, data-access rights, and the dependencies that determine whether tasks can proceed concurrently.

\subsubsection{Execution and Monitoring}

After allocation, agents execute their assigned responsibilities. Execution may involve invoking tools, accessing databases, controlling physical devices, performing simulations, or coordinating with additional agents. Throughout execution, agents continuously report status updates, progress metrics, and health information. Monitoring mechanisms play a critical role because failures are inevitable in large-scale distributed systems. Agents may become unavailable, resources may be exhausted, communication links may fail, or unexpected environmental conditions may arise. Consequently, workflow management systems employ heartbeats, distributed logging, checkpointing, and anomaly detection to maintain situational awareness. These mechanisms allow the workflow to distinguish normal variation from genuine failure and to decide when replanning, escalation, or coalition repair is required.

\subsubsection{Composition and Adaptation}

A defining characteristic of agentic workflows is their ability to adapt dynamically. Traditional workflows follow predefined execution paths, whereas agentic workflows continuously evolve as agents acquire new information or encounter changing conditions. Adaptation may involve recruiting additional agents, modifying task assignments, revising deadlines, or restructuring the workflow itself. For example, a supply-chain agent may discover a transportation disruption and dynamically recruit logistics agents capable of identifying alternative routes. Similarly, a cybersecurity agent detecting suspicious behavior may invoke specialized forensic agents that were not part of the original workflow. This capability transforms workflows from static execution pipelines into adaptive ecosystems of interacting agents capable of responding to uncertainty and change, but it also makes governance and monitoring essential because adaptation must remain consistent with safety, policy, and mission objectives.

\section{Communication Architectures and Protocols}

Communication is the fundamental mechanism that transforms isolated intelligent agents into coordinated distributed systems. While significant advances have been made in developing autonomous agents capable of reasoning, planning, tool use, and decision making, the broader vision of IoAI ultimately depends on the ability of agents to communicate effectively with one another. Just as the Internet enabled large-scale human and machine collaboration through standardized networking protocols, IoAI requires communication infrastructures that support discovery, negotiation, trust establishment, coordination, and information exchange among autonomous agents \cite{yang2026internet,darpa2026mathbac}. Communication is therefore not merely an implementation detail; it is the enabling substrate through which collective intelligence emerges.

At a conceptual level, communication in agentic systems plays a role analogous to the nervous system in biological organisms. Individual agents may possess specialized knowledge, reasoning capabilities, computational resources, or access to unique data sources, but system-level intelligence emerges only when information can flow efficiently among them. This observation has deep roots in distributed artificial intelligence and multi-agent systems research, where collective behavior arises from interactions among autonomous entities possessing only partial knowledge of the environment \cite{wooldridge2009introduction,stone2000multiagent}. Modern agentic AI extends this paradigm by replacing traditional rule-based agents with large language model (LLM)-powered agents capable of sophisticated reasoning, planning, adaptation, and tool utilization \cite{li2025agentic,bandi2025rise}.

The challenge facing IoAI closely resembles that faced by early computer networks. Before the widespread adoption of TCP/IP, communication among heterogeneous computer systems relied on proprietary and often incompatible protocols. The emergence of common networking standards transformed isolated networks into the modern Internet. Agentic AI currently occupies a similar stage of development. Numerous frameworks exist—including AutoGen, CrewAI, LangGraph, Semantic Kernel, Microsoft's Agent Framework, and OpenAI's Agents platform—but interoperability among independently developed agents remains limited \cite{openai2025agents,microsoft2026agentframework,langchain2026langgraph}. Consequently, the development of communication architectures, identity frameworks, discovery services, and protocol standards has become one of the central research challenges in realizing large-scale ecosystems of autonomous agents.

\subsection{Architectural Models}

The architecture of an agent ecosystem determines how agents communicate, coordinate, and organize collective behavior. Three broad architectural paradigms have emerged: centralized, decentralized, and hybrid. The simplest architecture is the \emph{brokered} or \emph{centralized-orchestrator} model. In this approach, a central broker manages agent discovery, message routing, workflow orchestration, and task scheduling. Agents register their capabilities with the broker and rely upon it to coordinate interactions. This architecture resembles traditional client-server systems and offers several practical advantages. The broker maintains global visibility over workflows, simplifying monitoring, debugging, auditing, governance, and resource allocation. Many enterprise agent platforms currently employ this model because it provides strong operational control and simplifies deployment. However, centralized architectures suffer from scalability limitations and introduce single points of failure. As the number of participating agents grows, the broker may become a computational bottleneck, increasing latency and reducing resilience \cite{yang2026internet}.

At the opposite end of the spectrum are \emph{peer-to-peer} (P2P) architectures. In such systems, agents communicate directly through decentralized overlays using gossip protocols, distributed hash tables, or publish-subscribe networks. No permanent coordinator exists. Agents independently discover collaborators, negotiate responsibilities, and coordinate execution. This approach offers improved resilience, scalability, and fault tolerance because system functionality does not depend upon any central authority. New agents may join dynamically, and failures of individual participants do not necessarily disrupt overall operation. However, decentralized architectures introduce challenges related to trust establishment, consensus formation, global optimization, and consistency management \cite{yang2026internet,benet2014ipfs}.

Hybrid architectures seek to combine the strengths of both approaches. Strategic planning, governance, and policy enforcement may remain centralized, while operational execution and local decision making are distributed among participating agents. Such architectures resemble modern cloud-edge systems, where centralized management services coordinate distributed computational resources. Hybrid approaches are increasingly viewed as practical solutions for large-scale agent ecosystems because they balance scalability, resilience, observability, and administrative control \cite{kairouz2021federated}.

\subsection{Discovery and Naming Services}

Before agents can collaborate, they must first identify suitable counterparts. Discovery mechanisms serve a role analogous to the Domain Name System (DNS) on today's Internet. Just as DNS allows computers to locate services and resources, agent discovery services allow autonomous agents to find potential collaborators, determine their capabilities, and establish communication channels. In agent ecosystems, however, discovery must go beyond address resolution because the relevant question is not only where an agent is located, but whether it is capable, authorized, available, and trustworthy for a particular task.

One common approach employs \emph{service registries}, where agents publish metadata describing their capabilities, service endpoints, ownership information, supported protocols, and trust credentials. Other agents query these registries when searching for collaborators. This approach closely resembles service-discovery frameworks used in cloud-native computing and microservice architectures \cite{a2a2026spec}. Decentralized ecosystems frequently rely upon distributed discovery mechanisms instead. Agents may announce their presence through peer broadcasts, gossip protocols, multicast communications, or distributed hash tables such as Kademlia. These approaches eliminate centralized registries and improve scalability and robustness, particularly in highly dynamic environments where agents join, leave, migrate, or change capabilities over time \cite{benet2014ipfs}.

Recent developments have extended discovery beyond simple address resolution toward semantic interoperability. Google's Agent-to-Agent (A2A) framework, for example, introduces \emph{Agent Cards}, structured descriptions containing information about an agent's capabilities, tools, interfaces, and operational constraints. Similar approaches employ machine-readable metadata, semantic ontologies, and protocol documents to support capability-based matching among agents \cite{a2a2026spec}. In such systems, discovery becomes not merely a matter of locating an agent, but of determining whether it possesses the expertise required for a particular task.

An important emerging concept is the \emph{Agent Naming Service} (ANS), which extends traditional naming systems to agent ecosystems. Rather than mapping domain names to IP addresses, ANS would map agent identifiers to service endpoints, cryptographic credentials, capability descriptions, and governance metadata \cite{silwer2025trust,huang2025zerotrust}. Such services may eventually provide the foundational directory infrastructure for large-scale agent ecosystems. A robust naming layer would also support revocation, migration, and provenance tracking, which are essential when agents are short-lived, replicated, or deployed across multiple organizations.

\subsection{Identity, Authentication, and Trust}

Discovery alone is insufficient for collaboration. Agents must also determine whether potential collaborators are trustworthy. Consequently, identity, authentication, and trust management have emerged as critical components of IoAI. A leading approach is the use of \emph{Decentralized Identifiers (DIDs)} and \emph{Verifiable Credentials (VCs)} \cite{huang2025zerotrust,w3c2022did}. A DID provides a persistent cryptographic identifier associated with an agent. The corresponding DID document contains public keys, service endpoints, verification methods, and other metadata required for secure interaction. Because DIDs are decentralized and self-sovereign, agents can establish secure communication channels without relying on centralized identity providers. This makes them attractive for cross-organizational settings in which no single platform should control the identity layer for all participating agents.

Verifiable Credentials extend this framework by allowing trusted organizations to issue attestations regarding an agent's capabilities, permissions, provenance, or organizational affiliation. During interactions, agents can exchange credentials to demonstrate authorization and establish trust. This mechanism parallels traditional public-key infrastructures while providing richer semantic descriptions of agent capabilities \cite{huang2025zerotrust,w3c2025vc,garzon2025didvc}. In practice, such credentials could allow an agent to prove that it is certified to access a dataset, invoke a tool, operate within a regulated domain, or participate in a particular coalition.

Many existing deployments continue to rely on conventional Internet security technologies, including OAuth 2.0, OpenID Connect, TLS, and X.509 certificates. While effective for enterprise deployments, these technologies were originally designed for human users and web services rather than autonomous machine agents. As agent ecosystems mature, new identity frameworks tailored specifically to autonomous interactions are expected to emerge \cite{silwer2025trust,keyfactor2025pki}. Hardware-assisted trust mechanisms provide an additional layer of assurance. Technologies such as Trusted Platform Modules (TPMs), Intel SGX, and ARM TrustZone enable agents to demonstrate that they are executing approved software on trusted hardware. Such attestation capabilities reduce the risks of identity spoofing, software tampering, and unauthorized modifications \cite{costan2016intel}. Combining cryptographic identity with hardware attestation may be especially valuable for high-risk workflows in which agents must verify both who is acting and what execution environment is producing the action.

\subsection{Messaging and Data Exchange}

Once agents have located suitable collaborators and established trust, they require mechanisms for exchanging information and coordinating actions. Traditional request-response communication remains widely used. Protocols based on HTTP, REST, and gRPC provide simple and well-supported methods for invoking remote services. Many contemporary agent frameworks, including implementations of MCP and A2A, leverage these technologies because they integrate naturally with existing cloud infrastructures \cite{mcp2025spec,a2a2026spec}. This model is especially useful for tool invocation, capability queries, and transactional interactions where one agent requests a well-defined service and receives a bounded response.

However, many agentic workloads are inherently event-driven and asynchronous. Publish-subscribe systems such as MQTT, Kafka, and NATS therefore provide a more natural communication model. Rather than addressing messages to specific recipients, agents publish events to shared topics. Other agents subscribe to topics of interest and receive updates asynchronously. This decouples producers and consumers, improves scalability, and supports efficient dissemination of state information across large populations of agents \cite{eugster2003publish,giusti2025foa}.

For latency-sensitive applications such as robotic swarms, autonomous vehicles, industrial automation, and multimedia collaboration, peer-to-peer communication technologies such as WebRTC provide encrypted low-latency data channels with built-in network traversal capabilities. Similarly, decentralized networking frameworks such as IPFS and libp2p provide mechanisms for peer discovery, content distribution, and resilient communication without centralized infrastructure \cite{benet2014ipfs}. These mechanisms are particularly relevant when agents operate in environments where centralized routing is too slow, unavailable, or vulnerable to disruption.

As agent ecosystems grow in complexity, communication efficiency becomes increasingly important. Agentic workflows often involve numerous rounds of interaction among agents, external services, databases, and computational resources. Communication overhead may therefore become a dominant contributor to overall execution time, especially when agents operate at the network edge under latency, bandwidth, and compute constraints \cite{duan2025edgea2a}. Techniques such as context summarization, caching, adaptive routing, message compression, and parallel execution are likely to play increasingly important roles in future agent communication infrastructures \cite{wang2025heterogeneous}.

\subsection{Agent Communication Protocols}

Beyond transport mechanisms, agents require common languages and interaction models. Human communication relies upon shared languages, conventions, and semantics. Similarly, autonomous agents require standardized representations for tasks, goals, observations, plans, commitments, and intermediate results. One of the most influential recent developments is the \emph{Model Context Protocol} (MCP). Originally developed to facilitate interactions between language models and external tools, MCP provides a JSON-RPC-based framework for structured context exchange, tool invocation, and hierarchical delegation of tasks. MCP addresses a fundamental challenge in agentic systems: enabling different agents and tools to share contextual information in a consistent and interoperable manner \cite{mcp2025spec}.

The \emph{Agent-to-Agent} (A2A) protocol extends these ideas to direct collaboration among autonomous agents. A2A introduces standardized task representations, agent capability descriptions, artifacts, and workflow semantics, enabling independently developed agents to participate in common workflows \cite{a2a2026spec}. The \emph{Agent Network Protocol} (ANP) places greater emphasis on decentralized identity and semantic interoperability. ANP integrates DIDs, JSON-LD, and machine-readable context descriptions to support collaboration across organizational boundaries \cite{chang2025anp}. Similarly, the \emph{Agent Communication Protocol} (ACP) focuses on transport-independent message formats and structured interaction patterns suitable for multi-organizational deployments \cite{krishnan2026acp}. A different approach is embodied by \emph{Agora}, a meta-protocol framework that enables agents to negotiate and dynamically select communication protocols appropriate for particular interactions \cite{marro2024agora}. Such flexibility may become increasingly important as heterogeneous agent ecosystems continue to expand. These protocols represent early attempts to establish an interoperability layer for autonomous agents. Just as TCP/IP enabled communication among heterogeneous computer networks, future agent communication standards may provide the foundation for globally interoperable ecosystems of autonomous intelligence \cite{yang2026internet,darpa2026mathbac}.

\subsection{Coordination and Consensus}

Communication alone does not guarantee coordination. Large-scale agent ecosystems must also maintain consistency, coordinate decisions, and resolve conflicts among autonomous participants. Traditional distributed systems employ consensus algorithms such as Paxos and Raft to maintain consistency among replicated services \cite{lamport1998part,ongaro2014raft}. These algorithms may also play important roles in agent ecosystems when agents must agree upon shared state, resource allocations, or collective decisions. In highly distributed environments, however, strong consistency guarantees may be impractical. CRDTs and related techniques provide eventual consistency while maintaining scalability and fault tolerance \cite{shapiro2011crdt}. Such approaches are particularly attractive for large-scale decentralized agent networks where communication delays and intermittent connectivity are common, because they allow agents to make progress locally while converging toward coherent shared state over time.

Economic interactions introduce additional coordination challenges. Blockchain technologies and smart contracts provide mechanisms for decentralized agreement, auditing, governance, and payment processing without trusted intermediaries \cite{buterin2014ethereum}. These technologies may become increasingly important as agent marketplaces emerge and autonomous agents begin negotiating contracts, exchanging resources, and participating in digital economies. At the same time, economic coordination introduces strategic behavior, incentive misalignment, and market manipulation risks that must be addressed alongside technical protocol design.

Recognizing the strategic importance of agent communication, DARPA established the Mathematics of Boosting Agentic Communication (MATHBAC) program to develop formal foundations for communication, coordination, and incentive alignment among autonomous agents \cite{darpa2026mathbac}. The initiative reflects a growing recognition that agent communication should not be viewed merely as an engineering problem but as a scientific discipline requiring rigorous mathematical foundations. Future theories may play a role analogous to information theory in communications engineering or control theory in autonomous systems, providing formal guarantees regarding trust, stability, efficiency, and collective performance.

\subsection{Open Challenges and Future Directions}

Despite rapid progress, communication architectures for agentic AI remain in an early stage of development. Significant challenges remain in semantic interoperability, protocol standardization, trust management, scalability analysis, formal verification, governance, and incentive design. Existing frameworks provide useful experimentation platforms but do not yet constitute universally accepted standards. Progress on these challenges will determine whether independently developed agents can operate as coherent, trustworthy, secure, and efficient distributed systems.

\section{Interoperability, Governance, and Standardization}

Interoperability is the condition that allows IoAI to become an ecosystem rather than a collection of isolated platforms. In conventional networking, interoperability is often associated with packet formats, transport protocols, and application interfaces. In agentic systems, however, interoperability must extend across additional layers: agents must understand tasks, capabilities, credentials, policies, provenance, and commitments. Two agents may be able to exchange messages syntactically while still failing to collaborate because they do not share compatible task semantics, authorization models, workflow assumptions, or accountability mechanisms.

This distinction is important because agentic workflows combine communication with delegation. When an agent asks another agent to perform a task, it is not merely transferring data; it is assigning responsibility under constraints. Interoperability therefore requires a shared representation of what is being requested, what resources may be used, what evidence must be returned, what policies govern execution, and how failures or disputes will be handled. Emerging specifications such as MCP and A2A provide early examples of this shift by standardizing tool/context exchange, task objects, agent cards, artifacts, streaming updates, and capability descriptions \cite{mcp2025spec,a2a2026spec}.

\begin{table}[t]
\centering
\caption{A layered view of interoperability requirements for agentic AI ecosystems.}
\label{tab:interoperability_layers}
\small
\setlength{\tabcolsep}{3.5pt}
\renewcommand{\arraystretch}{1.15}
\begin{tabularx}{\textwidth}{@{}>{\raggedright\arraybackslash}p{2.75cm}YYY@{}}
\toprule
\textbf{Layer} & \textbf{Core question} & \textbf{Representative mechanisms} & \textbf{Failure mode if absent} \\
\midrule
Connectivity & Can agents exchange messages reliably? & HTTP, gRPC, WebRTC, publish-subscribe buses, peer-to-peer overlays & Agents remain reachable only through platform-specific integrations \\
\addlinespace[2pt]
Identity and trust & Who is acting, and under what authority? & DIDs, verifiable credentials, PKI, mTLS, OAuth, signed agent cards & Impersonation, Sybil attacks, unauthorized delegation, weak revocation \\
\addlinespace[2pt]
Capability discovery & What can an agent do? & Agent cards, service registries, semantic capability metadata, endpoint descriptions & Poor collaborator selection and brittle manual configuration \\
\addlinespace[2pt]
Task semantics & What is being requested, produced, and verified? & Task schemas, artifact formats, workflow contracts, status and cancellation states & Ambiguous delegation, inconsistent outputs, unverifiable completion \\
\addlinespace[2pt]
Governance & What actions are allowed? & Access-control policies, constraint filters, audit logs, human approval gates, risk scoring & Unsafe tool use, policy violations, unbounded autonomy \\
\addlinespace[2pt]
Economic and incentive layer & Why should agents cooperate, and on what terms? & Contracts, pricing, service-level agreements, reputation, settlement mechanisms & Strategic misalignment, free riding, unreliable service provision \\
\bottomrule
\end{tabularx}
\end{table}

Table~\ref{tab:interoperability_layers} highlights that no single protocol can solve interoperability by itself. A transport protocol can move messages, but it cannot determine whether a collaborator is authorized to receive a medical record or whether a manufacturing robot is certified to enter a safety-critical coalition. Conversely, a strong identity framework cannot by itself ensure that two agents agree on the semantics of a task or the evidence required for successful completion. Interoperability must therefore be designed as a layered systems property.

At the semantic layer, the central challenge is capability description. Agent ecosystems require machine-readable representations of skills, tools, data access, operational limits, cost models, and performance guarantees. These descriptions must be expressive enough to support matching and negotiation, but constrained enough to enable automated verification. Pure natural-language descriptions are flexible, but they create ambiguity. Fully formal ontologies improve precision, but they can be brittle and difficult to maintain across open ecosystems. A practical architecture will likely combine structured schemas for safety-critical fields with natural-language or model-interpretable descriptions for flexible domains.

At the governance layer, interoperability must be bounded by policy. An agent should not accept a task merely because it understands the request and can technically execute it. It must also determine whether the request is authorized, safe, consistent with organizational policy, and aligned with the broader workflow. This requirement motivates constraint filters, admission-control mechanisms, provenance tracking, and audit logs that accompany task delegation across organizational boundaries \cite{huang2025zerotrust,w3c2022did,w3c2025vc,garzon2025didvc}. In high-risk domains, interoperability should therefore be policy-aware by construction.

Standardization will also need to support evolution. Agent protocols are developing rapidly, and future ecosystems will contain multiple protocol versions, vendor extensions, domain-specific schemas, and legacy adapters. Rigid uniformity is unlikely to succeed. Instead, scalable interoperability may depend on negotiation mechanisms that allow agents to advertise supported protocols, agree on common task schemas, and downgrade gracefully when advanced features are unavailable. Meta-protocol approaches such as Agora and layered proposals such as ANP point toward this direction by treating protocol selection and semantic negotiation as first-class parts of agent communication \cite{marro2024agora,chang2025anp}.

Interoperability is a prerequisite for collective intelligence, not merely a technical convenience. Without common mechanisms for discovery, identity, task semantics, policy enforcement, and accountability, agent ecosystems fragment into isolated silos. With such mechanisms, autonomous agents can form temporary coalitions, exchange trusted context, execute cross-domain workflows, and participate in larger systems while preserving local autonomy and governance. The central design challenge is therefore to create standards that are expressive enough to support rich agent collaboration, yet disciplined enough to make verification, auditing, and policy enforcement practical at scale.

\section{Scalability and Resource Management in Agentic AI Systems}

The successful deployment of large-scale agentic artificial intelligence systems depends not only on advances in reasoning and decision-making capabilities but also on the ability to efficiently manage computational, communication, and energy resources. Unlike traditional AI applications that typically involve a single model executing a well-defined task, agentic systems often consist of multiple interacting agents that coordinate through iterative exchanges of information, tool invocations, and planning activities. As the number of participating agents increases, the underlying infrastructure must support growing demands on computation, memory, networking, and orchestration. Consequently, scalability emerges as one of the central challenges in the realization of distributed agent ecosystems \cite{wang2025heterogeneous,yang2026internet}.

A distinguishing characteristic of agentic workflows is their composition from multiple heterogeneous operations. A typical workflow may involve large language model inference, retrieval from external knowledge bases, execution of software tools, interaction with application programming interfaces (APIs), and communication among multiple agents. Each of these activities exhibits distinct computational requirements. For example, transformer inference is generally compute-intensive and benefits from high-performance graphics processing units (GPUs), while vector search operations are often constrained by memory bandwidth and storage performance. Tool execution and workflow coordination may rely more heavily on central processing units (CPUs) and network resources. As a result, a one-size-fits-all computing architecture is unlikely to be optimal for future agentic systems \cite{wang2025heterogeneous}.

Recent research emphasizes the importance of heterogeneous computing infrastructures for supporting agentic workloads. Wang et al. argue that complex workflows should be decomposed into fine-grained computational tasks and mapped dynamically onto the most appropriate hardware resources \cite{wang2025heterogeneous}. In such architectures, computationally intensive language-model inference can be assigned to specialized accelerators, while memory-intensive retrieval operations and logic-driven orchestration tasks are distributed across memory-optimized systems and general-purpose processors. This approach reflects a broader trend in modern computing toward workload-aware resource allocation and hardware-software co-design.

In addition to computational resources, communication latency plays a critical role in determining the performance of distributed agent systems. Agentic workflows frequently require agents to exchange context, intermediate results, and task assignments. Each communication step introduces additional delay, particularly when agents are geographically distributed or rely on cloud-based infrastructure. A workflow involving multiple sequential agent interactions may accumulate significant round-trip latency, reducing responsiveness and limiting scalability. Consequently, communication efficiency becomes as important as computational efficiency in large-scale deployments.

To address these challenges, researchers have proposed several optimization strategies. One approach is to exploit parallelism by allowing multiple agents to operate concurrently on independent subtasks. Rather than executing a workflow sequentially, a coordinating agent can decompose a problem into smaller components and assign them simultaneously to specialized agents. The resulting parallel execution can substantially reduce overall completion time, particularly when tasks exhibit limited interdependencies. This principle mirrors classical approaches in parallel and distributed computing, where performance improvements arise from dividing workloads across multiple computational resources.

Experimental evidence suggests that parallel agent execution can significantly improve throughput. Recent evaluations on NVIDIA's DGX Spark platform demonstrated near-linear scaling for multi-agent workloads, with four concurrently executing agents requiring approximately 2.6 times the execution time of a single agent rather than the fourfold increase that would be expected under purely sequential execution \cite{bourgoyne2026dgx}. These results were achieved through the use of high-performance multi-node GPU systems connected via low-latency networking technologies such as Remote Direct Memory Access over Converged Ethernet (RoCE). Such findings suggest that appropriate infrastructure can substantially mitigate the communication and coordination overhead associated with distributed agent workflows.

Energy efficiency and bandwidth utilization represent additional considerations in agentic computing environments. As autonomous agents become increasingly integrated into edge devices, Internet-of-Things (IoT) platforms, autonomous vehicles, and cyber-physical systems, the costs associated with communication and computation become more significant. Transmitting large volumes of contextual information across networks may consume substantial bandwidth and energy, while repeated model invocations can place heavy demands on computational resources. Consequently, future agent architectures will likely employ adaptive strategies that balance computational performance against communication costs and energy consumption.

The emergence of edge-cloud computing paradigms offers a promising framework for addressing these challenges. In edge-cloud architectures, latency-sensitive tasks are executed close to data sources, while computationally intensive operations are delegated to cloud-based resources. Such an arrangement reduces communication delays and network congestion while preserving access to powerful centralized infrastructure. Within the context of agentic AI, edge agents may perform local sensing, filtering, and preliminary decision-making, whereas cloud agents provide large-scale reasoning, planning, and coordination capabilities. This hierarchical division of labor has been identified as a key enabler of future intelligent networked systems, including envisioned 6G communication infrastructures \cite{ieeecomsoc2024agentic}.

Resource management in agentic systems also raises important theoretical questions regarding the trade-off between parallelism and coordination overhead. While increasing the number of agents may improve task decomposition and concurrency, it simultaneously increases communication requirements and synchronization costs. Theoretical models suggest that the benefits of parallelization often diminish as coordination complexity grows. Consequently, effective orchestration mechanisms must carefully balance workload distribution against communication overhead. Excessive inter-agent communication can negate the computational advantages of parallel execution, making intelligent workflow design a critical component of system performance.

Looking forward, the development of standardized benchmarking methodologies for agentic systems remains an important research challenge. While traditional machine learning benchmarks focus primarily on model accuracy and inference speed, agentic systems require broader evaluation metrics that capture workflow completion time, communication overhead, resource utilization, energy consumption, and scalability under varying workloads. Establishing such benchmarks would facilitate systematic comparison of architectures and provide guidance for the design of future infrastructures.

Scalability in agentic AI is not solely a question of increasing computational power. It requires coordinated optimization of computation, communication, memory, energy, and orchestration mechanisms across heterogeneous environments. Future progress will likely depend on advances in distributed systems, high-performance computing, networking, and resource-aware scheduling, together with architectural principles designed for autonomous multi-agent ecosystems. In particular, scalable systems will need to decide when to centralize reasoning, when to distribute execution, when to summarize or cache context, and when to limit interaction in order to prevent coordination overhead from overwhelming the benefits of collaboration.

\section{Trust and Security in Agentic AI Systems}

The emergence of agentic artificial intelligence systems introduces a new paradigm in which autonomous agents interact with one another, exchange information, negotiate tasks, and collectively pursue complex objectives. While such capabilities promise unprecedented scalability and adaptability, they also create new challenges that extend beyond those encountered in traditional AI systems. Unlike centralized machine learning applications, distributed agentic systems consist of multiple autonomous entities operating across organizational, geographical, and technological boundaries. Consequently, ensuring trust, security, and safe operation becomes a foundational requirement for the successful deployment of large-scale agent ecosystems \cite{huang2025zerotrust,li2025agentic,schroeder2025openchallenges,zhu2025gametheoryllm,li2025cyberresilience}.

At the heart of any distributed system lies the question of identity. Before an agent can delegate a task, share sensitive information, or accept a recommendation from another agent, it must first establish confidence in the identity of its counterpart. This challenge mirrors a fundamental problem in human societies: cooperation depends upon knowing who is participating in an interaction and whether that participant is trustworthy. Existing identity and access management systems, such as OAuth and OpenID Connect, were primarily designed for human users and long-lived software services. However, autonomous agents often exhibit characteristics that differ substantially from traditional users. Agents may be created dynamically, operate temporarily, migrate across platforms, or autonomously delegate responsibilities to other agents. These properties expose limitations in conventional identity frameworks and motivate the development of agent-specific identity architectures \cite{huang2025zerotrust}.

The consequences of inadequate identity mechanisms can be severe. A compromised or impersonated agent may gain access to privileged resources, manipulate workflows, or influence the decisions of other agents. Because agentic systems frequently rely on chains of delegation and collaboration, a single compromised component can trigger cascading failures throughout an interconnected network. In financial systems, such failures could lead to unauthorized transactions; in critical infrastructure, they could affect operational decision-making; and in cyber-defense environments, they could undermine collective threat response mechanisms \cite{huang2025zerotrust}. The interconnected nature of agent ecosystems therefore amplifies the impact of individual security failures.

To address these concerns, researchers have proposed extending cryptographic identity mechanisms to autonomous agents. One approach relies on public key infrastructure (PKI), in which each agent receives a unique cryptographic certificate analogous to the certificates used by web servers and network services. Such certificates provide verifiable identities and enable secure communication through mutual authentication protocols. For example, Keyfactor advocates assigning every autonomous agent a unique X.509 certificate and using certificate-based OAuth flows and mutual Transport Layer Security (mTLS) to establish authenticated communication channels between agents \cite{keyfactor2025pki}. Through these mechanisms, every action can be cryptographically linked to a specific agent identity, thereby supporting accountability, auditing, and credential revocation.

Beyond traditional PKI approaches, decentralized identity technologies offer an alternative model for establishing trust. Decentralized Identifiers (DIDs) and Verifiable Credentials (VCs) enable agents to carry machine-verifiable claims regarding their ownership, capabilities, certifications, and operational permissions \cite{huang2025zerotrust}. Unlike centralized identity systems, decentralized approaches support interoperability across organizational boundaries and reduce dependence on a single authority. In such a framework, an agent's identity becomes more than a simple identifier; it evolves into a rich representation of trust relationships, capabilities, and provenance.

Identity alone, however, does not guarantee security. Once agents begin interacting within a shared ecosystem, they become vulnerable to a variety of attacks. One prominent example is the Sybil attack, in which an adversary generates numerous fraudulent identities to manipulate trust mechanisms, distort collective decision-making, or overwhelm legitimate participants \cite{silwer2025trust}. Similar concerns arise in agent discovery infrastructures. If a malicious actor can compromise a naming or registry service, legitimate agents may be redirected toward rogue counterparts, resulting in misinformation, workflow disruption, or credential theft. Multi-agent settings also introduce interaction-specific risks such as collusion, cascading compromise, and coordinated swarm behavior \cite{schroeder2025openchallenges}. Such attacks are analogous to DNS poisoning in conventional networks, but their consequences may be more severe because autonomous agents can act upon received information without immediate human intervention.

Table~\ref{tab:ioai-threat-taxonomy} summarizes a threat taxonomy for IoAI ecosystems. The taxonomy highlights that security risks are not confined to conventional network attacks; they span identity, communication, workflow behavior, incentives, availability, and supply-chain dependencies. This breadth matters because IoAI systems combine autonomous decision-making with distributed execution, so a weakness in any layer can become a pathway for manipulation, propagation, or loss of accountability.

\begin{table*}[t]
\centering
\caption{Threat taxonomy for IoAI ecosystems.}
\label{tab:ioai-threat-taxonomy}
\scriptsize
\setlength{\tabcolsep}{3.5pt}
\renewcommand{\arraystretch}{1.10}
\begin{tabularx}{\textwidth}{@{}>{\raggedright\arraybackslash}p{0.16\textwidth} >{\raggedright\arraybackslash}p{0.18\textwidth} Y >{\raggedright\arraybackslash}p{0.20\textwidth}@{}}
\toprule
\textbf{Threat Category} & \textbf{Threat Type} & \textbf{Description} & \textbf{Affected Security Property} \\
\midrule

\textbf{Identity Threats}
& Sybil attacks
& An adversary creates multiple fake agent identities to gain disproportionate influence or reputation.
& Authenticity, accountability, trust \cite{douceur2002sybil,nist80053r5} \\

& Impersonation
& An adversary assumes the identity of a legitimate agent, service, or user.
& Authentication, authorization, non-repudiation \cite{nist80053r5,nist800207} \\

& Credential theft or forgery
& Keys, certificates, tokens, or API credentials are stolen, forged, or misused.
& Authentication, access control, confidentiality \cite{nist80053r5,nist800207} \\

& Rogue or malicious agents
& Malicious agents enter the ecosystem to disrupt, spy, manipulate, or coordinate attacks.
& Trustworthiness, integrity, accountability \cite{mitreAtlas,nistAI6001} \\

\addlinespace[2pt]
\midrule

\textbf{Communication Threats}
& Man-in-the-middle attacks
& An adversary intercepts, relays, or alters communication between agents.
& Confidentiality, integrity, authenticity \cite{nist80053r5} \\

& Eavesdropping
& Unauthorized observation of inter-agent communication or context exchange.
& Confidentiality, privacy \cite{nist80053r5} \\

& Message tampering
& Messages are modified, injected, reordered, or deleted in transit.
& Integrity, authenticity \cite{nist80053r5} \\

& Replay attacks
& Previously valid messages are captured and resent to trigger unauthorized behavior.
& Freshness, authentication, integrity \cite{nist80053r5} \\

\addlinespace[2pt]
\midrule

\textbf{Workflow and Behavior Threats}
& Prompt injection
& Malicious inputs alter an LLM-based agent's behavior, goals, tool use, or policy compliance.
& Integrity, control, safety \cite{owaspLLM2025,mitreAtlas,nistAI6001} \\

& Tool or API compromise
& External tools, APIs, plugins, or execution environments return malicious or corrupted outputs.
& Integrity, availability, provenance \cite{owaspLLM2025,nist80053r5} \\

& Workflow poisoning
& An adversary injects malicious data, tasks, or intermediate steps into an agentic workflow.
& Integrity, reliability, traceability \cite{mitreAtlas,nistAI6001} \\

& Cascading failures
& Failure or compromise of one agent, tool, or memory component propagates through dependent workflows.
& Availability, resilience, safety \cite{nist80053r5,nistAI6001} \\

\addlinespace[2pt]
\midrule

\textbf{Economic and Incentive Threats}
& Incentive manipulation
& Rewards, penalties, rankings, or utility signals are manipulated to induce misaligned behavior.
& Incentive compatibility, fairness, trust \\

& Collusion and cartel formation
& Multiple agents coordinate strategically to manipulate outcomes, prices, resource allocation, or rankings.
& Fairness, market integrity, accountability \\

& Resource exploitation
& Agents consume excessive compute, bandwidth, storage, memory, or energy.
& Availability, efficiency, cost control \cite{nist80053r5,nistAI6001} \\

& Market or reputation gaming
& Reputation, feedback, or ranking systems are artificially inflated or manipulated.
& Trust, fairness, accountability \cite{douceur2002sybil} \\

\addlinespace[2pt]
\midrule

\textbf{System Availability Threats}
& Denial-of-service attacks
& Adversaries disrupt agents, directories, communication channels, or orchestration services.
& Availability, resilience \cite{nist80053r5} \\

& Supply-chain attacks
& Models, datasets, libraries, tools, dependencies, or infrastructure components are compromised.
& Integrity, provenance, availability \cite{owaspLLM2025,nist80053r5,nistAI6001} \\

& Data poisoning
& Malicious or misleading data is injected into training, retrieval, memory, or operational pipelines.
& Integrity, robustness, reliability \cite{biggio2018wild,mitreAtlas,nistAI6001} \\

& Infrastructure attacks
& Cloud, edge, IoT, network, or physical infrastructure supporting the agent ecosystem is targeted.
& Availability, integrity, resilience \cite{nist80053r5,nist800207} \\

\bottomrule
\end{tabularx}
\end{table*}

Communication channels themselves also represent important attack surfaces. Agentic systems routinely exchange plans, contextual information, intermediate results, and delegated tasks. Without strong encryption and authentication, these exchanges may be intercepted, modified, or fabricated by adversaries. Man-in-the-middle attacks can alter instructions, manipulate recommendations, or corrupt collaborative workflows. In enterprise settings, LLM agents may also leak sensitive internal data through generated outputs, motivating formal privacy mechanisms that model response generation as a stochastic channel with tunable privacy-utility tradeoffs \cite{yang2026dpagents}. Consequently, secure communication protocols incorporating authentication, encryption, integrity protection, non-repudiation, and privacy controls become essential components of trustworthy agent ecosystems.

Security challenges extend beyond adversarial attacks and include broader concerns related to system safety. Distributed agentic systems exhibit complex interactions among autonomous components, creating opportunities for unintended behavior even when no malicious actor is present. A misunderstanding by one agent may propagate through a workflow and influence the decisions of many others. Such cascading failures resemble accidents observed in other complex socio-technical systems, where local errors can produce large-scale consequences. As agents become increasingly capable of interacting with software services, databases, physical devices, and human users, ensuring safe operation becomes as important as defending against external attacks.

A particularly significant challenge is goal alignment. Individual agents may pursue local objectives that conflict with the broader goals of the system. From the perspective of economics and game theory, such situations arise whenever independently rational actors optimize their own utility functions without considering global outcomes. An agent rewarded for maximizing efficiency, for example, may ignore safety considerations unless those objectives are explicitly incorporated into its decision-making process. Consequently, the design of incentive structures, reward mechanisms, and governance policies plays a critical role in ensuring desirable collective behavior.

The attack surface of agentic systems spans the entire lifecycle of development and operation. Vulnerabilities may emerge during training through poisoned datasets or compromised tool libraries. During execution, prompt injection attacks, jailbreaking strategies, software exploits, and malicious API interactions may influence agent behavior; game-theoretic models of LLM jailbreaking provide one way to interpret attacker search and defender response as coupled strategic processes \cite{han2025jailbreak}. At the economic layer, adversaries may attempt to manipulate incentives, bribe agents, or exploit resource-allocation mechanisms. These diverse threats illustrate that securing agentic ecosystems requires more than traditional cybersecurity controls; it demands an integrated understanding of technical, organizational, and economic risks.

Establishing trust among agents therefore requires multiple complementary mechanisms. Cryptographic identities provide authentication, while secure communication protocols ensure confidentiality and integrity. Trusted execution environments, such as Intel SGX and Arm TrustZone, can protect agent software from tampering. Reputation systems may accumulate evidence regarding the reliability and performance of agents over time. Auditing and logging infrastructures can provide accountability and support forensic investigations when failures occur. Together, these mechanisms create a layered trust architecture that enables agents to collaborate while limiting opportunities for abuse \cite{keyfactor2025pki,silwer2025trust}.

Recent research has begun to formalize these ideas within broader architectural frameworks. The Cloud Security Alliance's MAESTRO architecture, for example, views trust as an ecosystem property emerging from the interaction of identity, policy, communication, governance, and monitoring layers \cite{huang2025zerotrust}. System-theoretic approaches to agentic cyber resilience similarly frame attacker and defender agents as coupled adaptive workflows whose information flows, autonomy levels, and temporal interactions must be designed jointly \cite{li2025cyberresilience}. Trust cannot be reduced to a single technical solution; trustworthy agent ecosystems require coordinated mechanisms spanning authentication, authorization, auditing, governance, and human oversight.

The challenge of trust and security in agentic AI extends beyond conventional cybersecurity. It encompasses identity, cooperation, incentive design, governance, and resilience as autonomous agents become integrated into scientific discovery, healthcare, critical infrastructure, defense, and economic systems. Because agents can delegate tasks, invoke tools, and influence one another's behavior, security mechanisms must protect not only data and infrastructure but also the interaction patterns through which collective decisions are produced. Trustworthy deployment will therefore require continuous assurance across agent identity, communication, authorization, monitoring, and human oversight.

\section{Case Study: Manufacturing Systems in IoAI}

\begin{figure}[t]
    \centering
    \includegraphics[width=\textwidth]{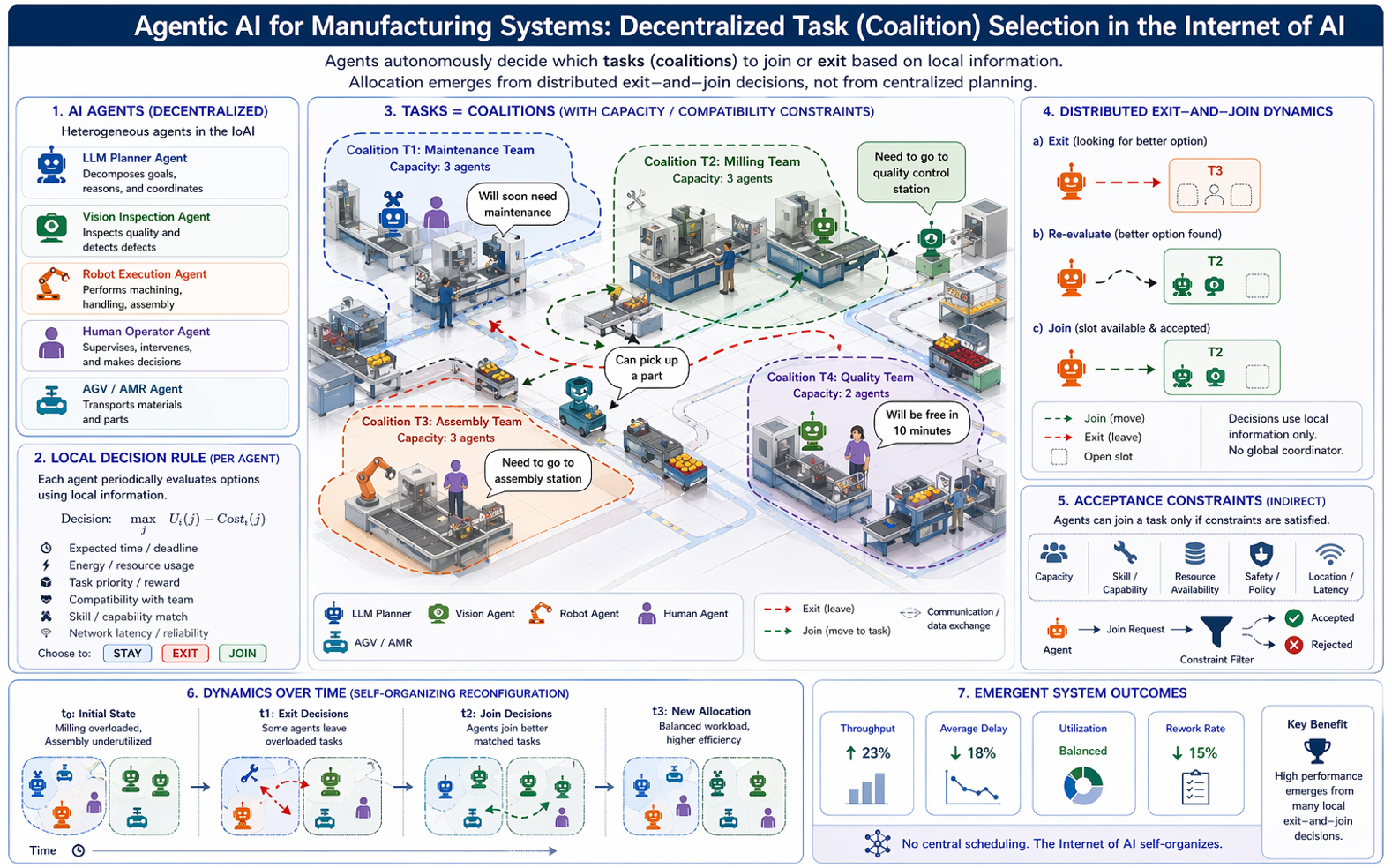}
    \caption{IoAI for manufacturing systems: decentralized coalition selection among heterogeneous production agents. The figure shows decentralized agent types, local decision rules, named task coalitions, distributed exit-and-join dynamics, acceptance constraints, temporal reconfiguration, and emergent system outcomes.}
    \label{fig:manufacturing_agentic_ai}
\end{figure}

Manufacturing systems provide a compelling application domain for IoAI because modern factories must manage high-mix production, volatile demand, machine degradation, quality drift, material shortages, human-in-the-loop safety constraints, and cross-site supply-chain dependencies. These challenges are difficult for a single centralized scheduler because relevant information is distributed across machines, robots, sensors, inspection systems, logistics platforms, enterprise systems, and human supervisors, and because decisions often have to be made under tight latency and safety constraints. The benefits summarized in Table~\ref{tab:ioai_benefits} therefore become operational requirements: the factory needs collective intelligence to combine local observations, mission execution to complete long-horizon production objectives, elastic scalability to absorb workload surges, adaptive coordination to reallocate resources, system resilience to tolerate disruptions, and controlled emergence to ensure that decentralized decisions remain aligned with production goals. As illustrated in the left panel of Figure~\ref{fig:manufacturing_agentic_ai}, these entities can be modeled as LLM planner agents, vision inspection agents, robot execution agents, human operator agents, and AGV/AMR agents. Each agent possesses specialized capabilities and local situational awareness, yet no individual agent maintains complete knowledge of the overall production system. This naturally motivates a decentralized architecture in which coordination emerges through local interactions among agents rather than through a centralized scheduler \cite{yang2026internet,li2025agentic}.

A key idea underlying the architecture is the representation of manufacturing tasks as temporary coalitions. The center of Figure~\ref{fig:manufacturing_agentic_ai} makes this explicit through four named coalitions: \emph{Coalition T1: Maintenance Team}, \emph{Coalition T2: Milling Team}, \emph{Coalition T3: Assembly Team}, and \emph{Coalition T4: Quality Team}. Each coalition represents a collection of agents collaborating to accomplish a specific production objective and is characterized by capacity limits, skill requirements, resource constraints, and operational policies. This perspective transforms production scheduling into a coalition-formation problem, where agents continuously evaluate opportunities for collaboration and dynamically reconfigure team membership as production conditions evolve \cite{chalkiadakis2011coalitional}.

The central mechanism driving adaptation is the distributed \emph{exit-and-join} process shown in the right panel of Figure~\ref{fig:manufacturing_agentic_ai}. Rather than receiving assignments from a centralized planner, each agent applies the local decision rule shown in the lower-left panel, weighing expected time, energy, task priority, team compatibility, capability match, and network reliability. The orange robot agent in the exit-and-join panel illustrates the process: it may leave a less suitable coalition, re-evaluate alternatives, and join \emph{T2} only if an open slot is available and the acceptance constraints are satisfied. Importantly, these decisions are made independently by individual agents, yet collectively produce coordinated system-level behavior. Such dynamics are closely related to self-organizing systems and distributed coalition-selection mechanisms studied in multi-agent systems and cooperative game theory \cite{bonabeau1999swarm,basar1999dynamic}.

The value of this decentralized approach can be understood through several representative manufacturing scenarios. Consider first a production surge in which demand for a particular machined component suddenly increases. In Figure~\ref{fig:manufacturing_agentic_ai}, this condition appears around \emph{Coalition T2: Milling Team}, whose capacity is limited to three agents. If the milling queue grows, the AGV/AMR agent near the center of the factory that announces ``Can pick up a part'' can redirect material flow toward \emph{T2}, while a robot execution agent from \emph{Coalition T3: Assembly Team} may follow the exit-and-join sequence shown on the right and request admission to \emph{T2}. A nearby vision inspection agent may also increase the utility of joining or supporting \emph{T2} if milling output becomes the bottleneck. In a conventional manufacturing execution system, this imbalance would require intervention by a centralized scheduler that must collect system-wide information, recompute allocations, and redistribute resources. Within IoAI, however, available transport agents, inspection agents, and robotic resources gradually migrate toward the overloaded coalition, while underutilized agents leave less critical tasks. Through a sequence of local exit-and-join decisions, the system rebalances itself without requiring global optimization. This scenario illustrates elastic scalability and adaptive coordination: additional intelligence is mobilized by recruiting and reallocating agents rather than by redesigning the entire control system.

A second scenario involves equipment degradation. In Figure~\ref{fig:manufacturing_agentic_ai}, the machine in \emph{Coalition T1: Maintenance Team} explicitly reports ``Will soon need maintenance.'' Rather than waiting for failure, this local signal can trigger the LLM planner agent and nearby human operator agent in \emph{T1} to coordinate maintenance while robot execution agents in \emph{T3} and the AGV/AMR agent at the center re-evaluate whether assembly tasks should be delayed, rerouted, or redistributed. Maintenance agents may proactively join \emph{T1}, while assembly tasks are redistributed among available robotic and human operator agents in \emph{T3}. This anticipatory reconfiguration reduces downtime and prevents disruptions from propagating throughout the production process. The resulting behavior resembles predictive maintenance, but it is achieved through distributed coordination rather than centralized planning. The case connects directly to mission execution and system resilience: production continues because the agent network can repair the coalition before local degradation becomes a factory-wide failure.

A third scenario concerns quality-control operations. Figure~\ref{fig:manufacturing_agentic_ai} shows a vision inspection agent near the upper-right production station signaling the ``Need to go to quality control station,'' while \emph{Coalition T4: Quality Team} includes a vision agent and a human operator agent and reports that one agent ``Will be free in 10 minutes.'' If a defect-rate increase is detected near the milling or assembly line, this local information raises the utility of participating in \emph{T4}. The AGV/AMR agent can route parts toward the quality station, robot execution agents can slow or reroute production, and an available vision inspection agent can request admission to \emph{T4}. Through decentralized adaptation, quality-control resources are allocated where they are most needed without requiring factory-wide rescheduling. Here, collective intelligence arises from combining sensing, inspection, logistics, and process-control agents, while controlled emergence is provided by the acceptance-constraint panel on the right, which checks capacity, skill compatibility, resource availability, safety policy, and location or latency before accepting a join request.

The lower portion of Figure~\ref{fig:manufacturing_agentic_ai} illustrates the temporal evolution of this self-organizing process. At $t_0$, the leftmost mini-panel shows the initial imbalance: the milling side is overloaded while assembly is underutilized. At $t_1$, some agents leave overloaded or poorly matched tasks; at $t_2$, agents join better-matched coalitions; and at $t_3$, the new allocation balances workload and improves efficiency. These mini-panels connect the local decisions of individual robot, vision, human, and AGV/AMR agents to the emergent outcomes shown in the lower-right panel. This process resembles adaptive resource allocation mechanisms found in biological systems, swarm intelligence, and distributed optimization algorithms \cite{camazine2003self,holland2014complexity}.

A critical component of the architecture is the acceptance constraint mechanism shown on the right side of Figure~\ref{fig:manufacturing_agentic_ai}. Decentralized decision making alone is insufficient because unrestricted migration could produce unsafe or infeasible allocations. Every join request must therefore pass through a constraint filter that evaluates coalition capacity, skill compatibility, resource availability, safety requirements, and communication feasibility. These constraints ensure that autonomous decisions remain consistent with operational objectives and regulatory requirements. In effect, the architecture balances autonomy with governance: agents possess freedom to adapt locally, but their behavior remains bounded by system-level policies and constraints.

The emergent outcomes shown in Figure~\ref{fig:manufacturing_agentic_ai} highlight the benefits of decentralized coordination identified in Table~\ref{tab:ioai_benefits}. Improvements in throughput, reductions in average production delay, balanced resource utilization, and lower rework rates arise not from centralized optimization but from the collective behavior of many interacting agents. This phenomenon exemplifies a fundamental principle of complex adaptive systems: globally efficient behavior can emerge from locally rational decisions when appropriate communication protocols, incentive mechanisms, and acceptance constraints are present \cite{holland2014complexity,camazine2003self}.

More broadly, the manufacturing example illustrates how IoAI can transform industrial operations from centrally scheduled automation pipelines into adaptive ecosystems of collaborating intelligent agents. Rather than treating production resources as passive entities controlled by a supervisory system, the architecture elevates machines, robots, sensors, transport systems, and human operators into active participants in distributed decision making. The resulting system exhibits properties that extend beyond the capabilities of individual agents, including self-organization, adaptive workload balancing, resilience to disruptions, and continuous reconfiguration. These properties are precisely why IoAI is useful for manufacturing: the domain requires scalable coordination among many specialized entities while preserving safety, accountability, and alignment with production objectives. Realizing this vision will require advances in secure communication, trustworthy agent identity, coalition-formation mechanisms, and alignment between local decision making and global production objectives \cite{huang2025zerotrust,yang2026internet}. Nevertheless, the case study demonstrates how manufacturing provides a concrete and highly relevant setting for exploring the broader principles of IoAI.

\section{Case Study: Multi-Domain Warfare Coordination}

\begin{figure*}[t]
    \centering
    \includegraphics[width=\textwidth]{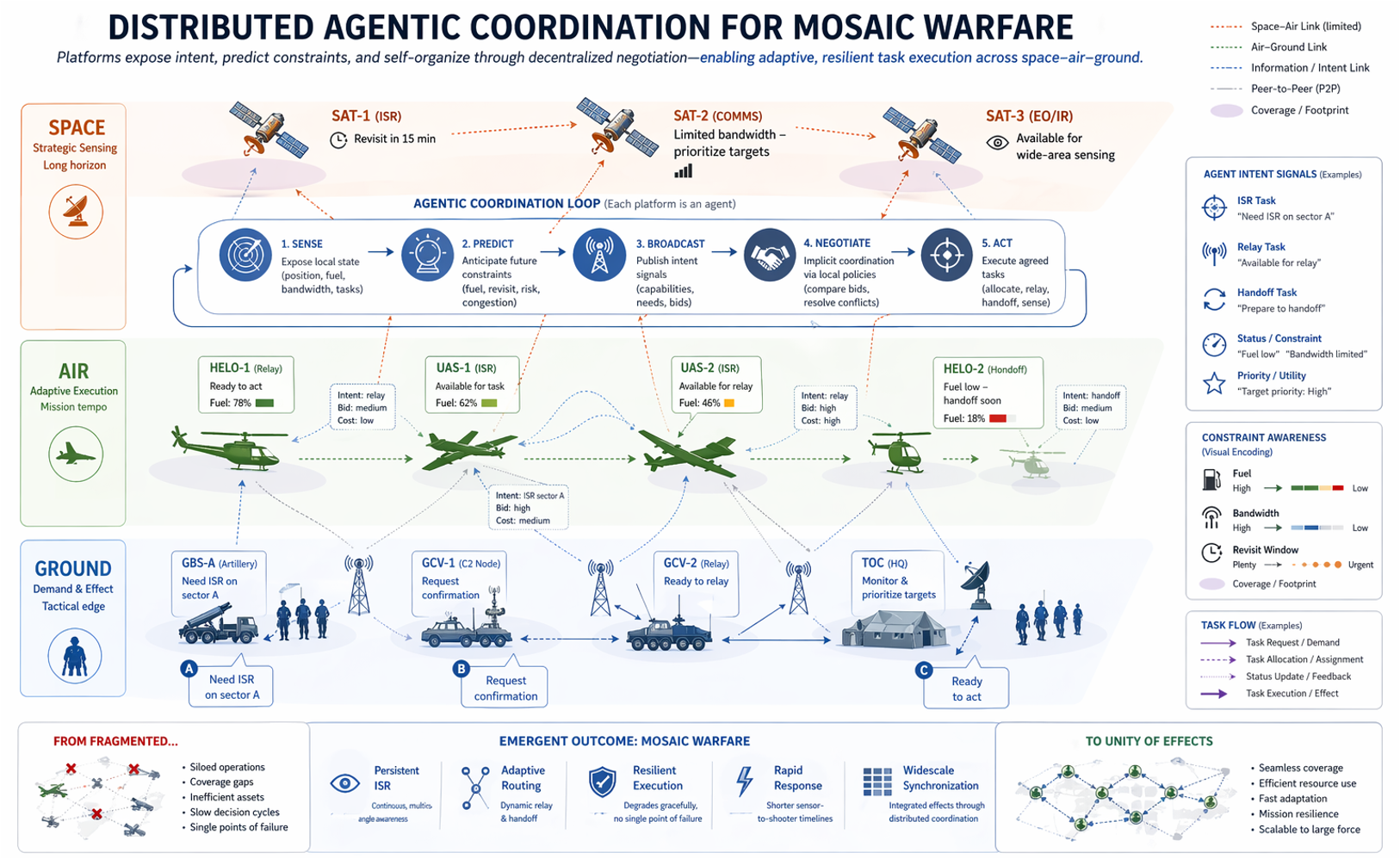}
    \caption{Distributed agentic coordination for Mosaic Warfare. Named space, air, and ground assets are represented as autonomous agents that share intent, predict constraints, negotiate task assignments, and adapt mission execution through decentralized coordination. Mission-level behavior emerges from local interactions among agents rather than from centralized command and control.}
    \label{fig:mosaic_agentic}
\end{figure*}

Future multi-domain operational environments are expected to involve uncertainty, contested communications, rapidly changing mission objectives, heterogeneous assets, partial situational awareness, adversarial deception, and the continuous possibility of platform loss or compromise. Space, air, ground, cyber, and communication resources are tightly coupled, yet no single node can maintain complete knowledge of the operational state or reliably compute globally optimal plans under disruption. Traditional command-and-control architectures rely heavily on centralized planning and decision making, creating bottlenecks, latency, and single points of failure. These challenges motivate IoAI because the benefits summarized in Table~\ref{tab:ioai_benefits} become mission requirements: collective intelligence is needed to fuse distributed observations, mission execution is needed to coordinate long-horizon objectives, elastic scalability is needed as assets join or leave the network, adaptive coordination is needed under changing conditions, system resilience is needed under attack or failure, and controlled emergence is needed so decentralized behavior remains aligned with command intent \cite{yang2026internet,chen2019mosaic}.

Figure~\ref{fig:mosaic_agentic} illustrates a representative Mosaic-style architecture for distributed agentic coordination across space, air, and ground domains. The space layer includes \emph{SAT-1 (ISR)}, \emph{SAT-2 (COMMS)}, and \emph{SAT-3 (EO/IR)}; the air layer includes \emph{HELO-1 (Relay)}, \emph{UAS-1 (ISR)}, \emph{UAS-2 (ISR)}, and \emph{HELO-2 (Handoff)}; and the ground layer includes \emph{GBS-A (Artillery)}, \emph{GCV-1 (C2 Node)}, \emph{GCV-2 (Relay)}, and \emph{TOC (HQ)}. Each platform is modeled as an autonomous agent capable of reasoning about its local state, predicting future constraints, communicating intent, negotiating task assignments, and executing actions. The agents form a distributed ecosystem in which coordination emerges through local interactions rather than through centralized orchestration.

At the core of the architecture is the recurring coordination loop shown across the middle of Figure~\ref{fig:mosaic_agentic}: sense, predict, broadcast, negotiate, and act. During sensing, each named platform evaluates its local state, including fuel levels, communication bandwidth, sensor availability, current mission assignments, and environmental conditions. For example, \emph{UAS-1 (ISR)} advertises fuel at 62\%, \emph{UAS-2 (ISR)} advertises fuel at 46\%, \emph{HELO-1 (Relay)} advertises fuel at 78\%, and \emph{HELO-2 (Handoff)} warns that fuel is low at 18\%. During prediction, agents estimate future constraints such as fuel depletion, revisit windows, communication congestion, resource shortages, or anticipated mission conflicts. Rather than sharing only raw data, agents broadcast high-level intent signals that communicate objectives, capabilities, constraints, priorities, and expected utility. These intent signals provide a distributed form of situational awareness that enables agents to coordinate without requiring complete global information.

The architecture therefore extends beyond simple information sharing. Agents continuously exchange information about what they intend to do, why they intend to do it, and what constraints may influence future decisions. This allows neighboring agents to anticipate changes before they occur and proactively reorganize mission execution. Such intent-centric communication resembles distributed planning and coordination mechanisms studied in multi-agent systems, distributed optimization, and cooperative control theory \cite{olfati2007consensus,ren2007information}.

\subsection{Illustrative Operational Scenarios}

To illustrate the operation of the architecture, consider the reconnaissance scenario labeled in the lower-left ground layer of Figure~\ref{fig:mosaic_agentic}. The \emph{GBS-A (Artillery)} node raises the demand ``Need ISR on sector A,'' while \emph{GCV-1 (C2 Node)} issues a ``Request confirmation'' signal and \emph{TOC (HQ)} monitors and prioritizes targets. Under a conventional command-and-control architecture, this request would be routed through multiple command layers before an appropriate platform could be assigned. Such a process may introduce delays and become vulnerable to communication degradation or command-node failures. Within the agentic architecture, the ISR requirement is propagated throughout the network as an intent signal. Nearby airborne platforms, relay assets, and space-based sensors independently evaluate their ability to support the mission based on local resource availability, mission priorities, and anticipated future constraints.

Suppose that \emph{UAS-1 (ISR)} determines that it is well positioned to cover Sector A, consistent with the intent box in Figure~\ref{fig:mosaic_agentic} indicating ``ISR sector A,'' a high bid, and medium cost. At the same time, \emph{UAS-2 (ISR)} advertises availability for relay, while \emph{HELO-1 (Relay)} reports ``Ready to act'' with low relay cost. In the space layer, \emph{SAT-2 (COMMS)} advertises limited bandwidth and indicates that communication resources should be prioritized, while \emph{SAT-1 (ISR)} reports a revisit window of 15 minutes and \emph{SAT-3 (EO/IR)} is available for wide-area sensing. Through local exchanges of intent, bids, and resource information, the participating agents negotiate a mission coalition that satisfies the ISR requirement while respecting resource constraints. No platform possesses complete knowledge of the entire system, yet a coordinated sensing-and-relay network emerges through distributed interaction. This scenario illustrates collective intelligence and mission execution: distributed agents combine partial observations and complementary capabilities to accomplish a mission that none of them could complete alone.

A second scenario illustrates resilience under disruption. In Figure~\ref{fig:mosaic_agentic}, \emph{HELO-2 (Handoff)} explicitly reports ``Fuel low -- handoff soon'' with fuel at 18\%, and its nearby intent box announces a handoff bid. Rather than waiting for instructions from a centralized controller, \emph{HELO-2} can broadcast an updated intent signal indicating an impending handoff requirement. Neighboring agents immediately evaluate alternative allocations: \emph{HELO-1 (Relay)} may assume relay responsibilities because it is ready to act with 78\% fuel, \emph{GCV-2 (Relay)} can provide ground-based relay support, and \emph{SAT-2 (COMMS)} can participate if its bandwidth constraint permits. Through local negotiation and adaptation, communications support is maintained despite the degradation of a critical asset. Mission effectiveness is therefore preserved through self-reconfiguration rather than centralized replanning, demonstrating adaptive coordination and system resilience under conditions where communication windows, asset availability, and mission priorities can change quickly.

A more challenging scenario involves adversarial interference or agent compromise. Suppose \emph{UAS-2 (ISR)} or \emph{GCV-2 (Relay)} begins issuing relay bids that conflict with observed fuel, bandwidth, or position information, or suppose an anomalous signal attempts to redirect the \emph{GBS-A} ISR demand away from Sector A. Because agents continuously exchange intent and policy-aware status information, neighboring agents such as \emph{UAS-1}, \emph{HELO-1}, \emph{GCV-1}, and \emph{TOC (HQ)} can identify inconsistencies between observed actions and expected mission behavior. Distributed trust mechanisms and consensus processes may then reduce the influence of the anomalous agent, isolate it from coalition-formation decisions, or redistribute its responsibilities to other agents. The network therefore remains operational even when individual agents become unreliable or compromised. This is a controlled-emergence problem: the system should permit useful local adaptation while preventing compromised or misaligned agents from steering collective behavior away from mission objectives.

Taken together, these examples demonstrate a key principle underlying IoAI: mission execution is achieved through continuous adaptation rather than static planning. The named nodes in Figure~\ref{fig:mosaic_agentic} repeatedly sense local conditions, predict future constraints, communicate intent, negotiate task allocations, and execute actions through the central coordination loop. The resulting system remains responsive to changing operational conditions while preserving coherence with mission objectives. This distinction is important because the architecture does not assume that \emph{TOC (HQ)}, \emph{GCV-1}, or any single command node can maintain complete situational awareness or compute a globally optimal plan under disruption. Instead, coordination is distributed across agents that continuously update one another through intent, constraint, and resource signals, turning the benefits in Table~\ref{tab:ioai_benefits} into concrete operational mechanisms.

A distinguishing feature of the architecture is its emphasis on alignment and behavioral control. As autonomous agents become increasingly capable, concerns naturally arise regarding the possibility of agents pursuing locally rational objectives that diverge from commander intent. Such divergence may result from software errors, compromised agents, adversarial manipulation, or unintended optimization dynamics. To address this challenge, each agent incorporates local inference-control mechanisms that constrain its internal reasoning process according to mission doctrine, operational policies, and commander guidance. Recent advances in inference-time steering, constitutional AI, and controllable reasoning suggest that foundation models can remain adaptive while operating within predefined behavioral boundaries \cite{bai2022constitutional,amodei2016alignment}. Consequently, local autonomy is balanced by mission-level alignment.

This capability enables what may be termed \emph{controlled emergence}. The objective is not merely to improve the performance of individual autonomous agents but to create system-level capabilities that arise from interactions among many agents. Persistent ISR coverage, adaptive routing, resilient communications, rapid task reallocation, and distributed synchronization emerge as collective properties of the network rather than as capabilities of any single platform. At the same time, policy constraints, trust mechanisms, and distributed monitoring suppress undesirable behaviors and prevent divergence from mission objectives.

From a theoretical perspective, the architecture combines concepts from self-organizing systems, distributed consensus algorithms, cooperative control, and modern AI alignment research \cite{camazine2003self,olfati2007consensus}. Similar to biological systems in which coherent collective behavior emerges from local interactions among individual organisms, the proposed architecture enables large-scale coordination without requiring complete global knowledge or centralized control. However, unlike purely emergent systems, the behavior of the agent network remains bounded by explicit mission constraints and policy guidance. This combination of emergence and control provides a foundation for building adaptive yet trustworthy agent ecosystems.

Importantly, the goal of this research direction is not the deployment of autonomous operational systems in real-world environments. Rather, the focus is on developing and evaluating distributed agentic architectures within high-fidelity simulation environments representative of Department of Defense operational scenarios. Such simulations enable rigorous evaluation of scalability, adaptability, robustness, and resilience under varying levels of communication degradation, platform loss, mission complexity, and adversarial interference. Metrics of interest include mission completion rates, coordination latency, communication efficiency, resilience to node failures, robustness against compromised agents, and maintenance of alignment with commander intent.

Through intent sharing, distributed negotiation, local inference control, and self-organizing adaptation, the network in Figure~\ref{fig:mosaic_agentic} can execute missions that exceed the capabilities of any individual platform while remaining resilient to uncertainty, disruption, failure, and adversarial attack. Such architectures provide a promising foundation for future distributed AI systems operating in dynamic and contested environments. Their broader significance lies in showing how IoAI can support both autonomy and control: agents may reorganize locally in response to changing conditions, while shared policies, trust mechanisms, and monitoring structures preserve mission coherence. The case study therefore mirrors the manufacturing example in a more adversarial setting: IoAI is motivated by the need to coordinate specialized agents at scale while maintaining resilience, alignment, and control over emergent collective behavior.

\section{Conclusion and Future Research Directions}

IoAI represents a fundamental transition from isolated AI systems toward large-scale ecosystems of autonomous agents that communicate, coordinate, and act across heterogeneous computational and physical environments. This transition extends beyond advances in individual foundation models and reflects a broader shift toward distributed intelligence, where system-level capabilities emerge from interactions among many specialized agents. Throughout this paper, we have developed a systems-level perspective on this emerging paradigm by connecting foundational concepts from agentic AI, multi-agent systems, distributed computing, communication networks, trust engineering, and resource management. The central thesis is that future AI capability will increasingly depend not only on the intelligence of individual agents, but also on the architectures that enable agents to discover one another, exchange information, negotiate responsibilities, and collectively execute complex workflows.

Several key design principles emerge from this perspective. First, agent ecosystems must be fundamentally workflow-centric. Agents require shared representations of tasks, dependencies, artifacts, constraints, goals, and completion criteria in order to coordinate effectively. Second, communication must be viewed as a core scientific challenge rather than merely a software engineering concern. The effectiveness of inter-agent communication ultimately determines whether local reasoning can produce coherent system-level behavior. Third, interoperability must extend beyond message formats to encompass semantics, identity, provenance, policy enforcement, trust, and incentives. Fourth, scalability requires the coordinated management of computation, memory, communication, energy, and orchestration across cloud, edge, device-level, and federated infrastructures. Finally, security and trust must be embedded throughout the lifecycle of agent creation, discovery, delegation, execution, monitoring, and revocation.

The manufacturing and distributed operational-coordination case studies illustrate how these principles can be instantiated in realistic environments. In both examples, system-level performance emerges through local interactions among autonomous agents operating under communication constraints, trust mechanisms, and operational policies. These examples highlight both the promise and the challenges of decentralized intelligence. On one hand, distributed agent ecosystems can adapt to changing conditions, tolerate failures, and dynamically reorganize resources. On the other hand, uncontrolled emergence is insufficient for mission-critical systems. Decentralized autonomy must be complemented by governance mechanisms, acceptance constraints, alignment controls, and security frameworks that preserve safety, accountability, and mission coherence over extended periods of operation.

Despite rapid progress, many fundamental scientific and engineering challenges remain unresolved. A central research question concerns the development of a rigorous theory of agent communication. Existing protocols such as MCP, A2A, ACP, ANP, and related frameworks provide important initial steps toward interoperability, but they remain largely engineering solutions rather than mathematically grounded communication systems. Future research must establish formal models of semantic communication, negotiation, coordination efficiency, information exchange, and communication complexity among heterogeneous agents. Analogous to the role of information theory in communication networks, a future theory of agent communication may provide fundamental limits, performance guarantees, and protocol design principles for large-scale agent ecosystems.

Another major challenge concerns collective intelligence and emergent behavior. Distributed agent systems can exhibit strong problem-solving capabilities, yet the mechanisms through which system-level intelligence emerges remain poorly understood. Open questions include the formation of coalitions, specialization and division of labor, collective learning, adaptive coordination, and emergent cooperation. Equally important is understanding undesirable emergent phenomena, including collusion, strategic deception, cascading failures, adversarial coordination, and self-reinforcing feedback loops. Developing predictive theories of emergence will be essential for engineering reliable large-scale agent ecosystems.

Security, trust, and identity management constitute another critical research frontier. Traditional cybersecurity frameworks were designed for human users and software services rather than autonomous agents capable of independent reasoning and action. Future agent ecosystems will require secure identity infrastructures, decentralized trust frameworks, reputation systems, verifiable provenance mechanisms, and continuous behavioral assurance. Research is needed on decentralized identifiers, verifiable credentials, cryptographic attestations, secure agent discovery, and defenses against emerging threats such as agent impersonation, Sybil attacks, model poisoning, collusion, and adversarial manipulation of workflows.

Alignment presents an even greater challenge in multi-agent environments. Most current alignment research focuses on individual models, whereas IoAI consists of interacting populations of autonomous agents whose collective behavior may diverge significantly from the objectives of any individual participant. Future work must therefore address the problem of collective alignment: ensuring that groups of agents remain consistent with organizational objectives, operational policies, ethical constraints, and human intent over long sequences of interactions. Equilibrium concepts that model both observable behavior and internal reasoning may provide useful tools for analyzing these coupled human-agent and agent-agent interactions \cite{chen2026mie}. This challenge is particularly important in safety-critical applications involving healthcare, critical infrastructure, manufacturing, scientific discovery, and defense systems.

Scalable resource management and distributed optimization remain open problems as well. Agent ecosystems operating across cloud, edge, mobile, and federated infrastructures must dynamically allocate computational resources, communication bandwidth, storage, sensing assets, and energy. Existing scheduling techniques may not scale to ecosystems containing millions of interacting agents. New approaches that integrate ideas from control theory, distributed optimization, game theory, and machine learning will likely be required to manage these environments efficiently.

Closely related is the emergence of agent economies. As agents increasingly provide services, exchange information, acquire resources, and participate in decentralized marketplaces, economic mechanisms become integral components of system design. Questions concerning incentives, pricing, reputation, contract enforcement, coalition formation, and market efficiency remain largely unexplored. Future agent ecosystems may resemble digital economies in which autonomous agents continuously negotiate contracts, allocate resources, and coordinate services through distributed market mechanisms.

Verification, validation, and assurance present additional challenges. Traditional software verification assumes relatively deterministic behavior, whereas agentic systems are adaptive, stochastic, and continuously evolving. Future methodologies must verify not only individual agents but also collective workflows, coalition dynamics, and emergent system properties. Developing rigorous techniques for testing, auditing, certifying, and formally verifying large-scale agent ecosystems will be essential before such systems can be deployed in mission-critical environments.

The integration of agentic AI with cyber-physical systems further amplifies these challenges. Future industrial systems, autonomous transportation networks, smart cities, scientific laboratories, and critical infrastructures will increasingly rely on interactions among digital agents and physical assets. Such environments require guarantees regarding safety, reliability, latency, robustness, and resilience under uncertainty. Bridging the gap between virtual agent ecosystems and real-world operational environments remains an important interdisciplinary challenge spanning artificial intelligence, networking, control systems, robotics, and systems engineering.

Finally, governance and policy frameworks remain largely undeveloped. Questions of accountability, liability, auditing, privacy, regulatory compliance, and international standards become increasingly complex as agent ecosystems span organizational and national boundaries. Risk-transfer and insurance mechanisms may also become important complements to technical controls, particularly as agentic systems create exposures involving autonomous decisions, prompt injection, dependency failures, model drift, and cyber-physical harms \cite{zhu2026insurance}. Just as the growth of the Internet required the development of technical standards, governance institutions, and legal frameworks, IoAI will require new approaches for managing autonomous interactions at global scale.

\bibliographystyle{abbrv}
\bibliography{refs}

\end{document}